\documentclass[%
 reprint,
 amsmath,amssymb,
superscriptaddress
]{revtex4-1}

\usepackage{graphicx}
\usepackage{dcolumn}
\usepackage{bm}

\usepackage{tikz}

\usepackage{lipsum}

\usepackage{physics}
\usepackage{hyperref}
\hypersetup{
    colorlinks=true,
    linkcolor=red,
    filecolor=magenta,      
    urlcolor=cyan,
}

\newcommand{\unit}[1]{\,\mathrm{#1}}

\begin{document}

\title{Hardware-Efficient Microwave-Activated Tunable Coupling Between Superconducting Qubits}

\author{Bradley K. Mitchell}
\thanks{These two authors contributed equally. Correspondence should be addressed to bradmitchell@berkeley.edu and rnaik24@berkeley.edu}
\affiliation{Quantum Nanoelectronics Laboratory, University of California, Berkeley, Berkeley CA 94720}
\affiliation{Computational Research Division, Lawrence Berkeley National Laboratory, Berkeley CA 94720}
\author{Ravi K. Naik}
\thanks{These two authors contributed equally. Correspondence should be addressed to bradmitchell@berkeley.edu and rnaik24@berkeley.edu}
\affiliation{Quantum Nanoelectronics Laboratory, University of California, Berkeley, Berkeley CA 94720}
\affiliation{Computational Research Division, Lawrence Berkeley National Laboratory, Berkeley CA 94720}
\author{Alexis Morvan}
\affiliation{Quantum Nanoelectronics Laboratory, University of California, Berkeley, Berkeley CA 94720}
\affiliation{Computational Research Division, Lawrence Berkeley National Laboratory, Berkeley CA 94720}
\author{Akel Hashim}
\affiliation{Quantum Nanoelectronics Laboratory, University of California, Berkeley, Berkeley CA 94720}
\affiliation{Computational Research Division, Lawrence Berkeley National Laboratory, Berkeley CA 94720}
\author{John Mark Kreikebaum}
\affiliation{Quantum Nanoelectronics Laboratory, University of California, Berkeley, Berkeley CA 94720}
\affiliation{Materials Science Division, Lawrence Berkeley National Laboratory, Berkeley CA 94720}
\author{Brian Marinelli}
\affiliation{Quantum Nanoelectronics Laboratory, University of California, Berkeley, Berkeley CA 94720}
\affiliation{Computational Research Division, Lawrence Berkeley National Laboratory, Berkeley CA 94720}
\author{Wim Lavrijsen}
\affiliation{Computational Research Division, Lawrence Berkeley National Laboratory, Berkeley CA 94720}
\author{Kasra Nowrouzi}
\affiliation{Quantum Nanoelectronics Laboratory, University of California, Berkeley, Berkeley CA 94720}
\affiliation{Computational Research Division, Lawrence Berkeley National Laboratory, Berkeley CA 94720}
\author{David I. Santiago}
\affiliation{Quantum Nanoelectronics Laboratory, University of California, Berkeley, Berkeley CA 94720}
\affiliation{Computational Research Division, Lawrence Berkeley National Laboratory, Berkeley CA 94720}
\author{Irfan Siddiqi}
\affiliation{Quantum Nanoelectronics Laboratory, University of California, Berkeley, Berkeley CA 94720}
\affiliation{Computational Research Division, Lawrence Berkeley National Laboratory, Berkeley CA 94720}
\affiliation{Materials Science Division, Lawrence Berkeley National Laboratory, Berkeley CA 94720}

\date{\today}

\begin{abstract}
  Generating high-fidelity, tunable entanglement between qubits is crucial for realizing gate-based quantum computation. In superconducting circuits, tunable interactions are often implemented using flux-tunable qubits or coupling elements, adding control complexity and noise sources. Here, we realize a tunable $ZZ$ interaction between two transmon qubits with fixed frequencies and fixed coupling, induced by driving both transmons off-resonantly. We show tunable coupling over one order of magnitude larger than the static coupling, and change the sign of the interaction, enabling cancellation of the idle coupling. Further, this interaction is amenable to large quantum processors: the drive frequency can be flexibly chosen to avoid spurious transitions, and because both transmons are driven, it is resilient to microwave crosstalk. We apply this interaction to implement a controlled phase (CZ) gate with a gate fidelity of $99.43(1)\%$ as measured by cycle benchmarking, and we find the fidelity is limited by incoherent errors.
\end{abstract}

\maketitle

High-fidelity two-qubit entangling operations, or two-qubit gates, are critical for many proposals for building quantum computers that can outperform classical computers at specific tasks~\cite{kjaergaard_superconducting_2020}. Important qualities for entangling schemes include high fidelity, low leakage~\cite{ghosh_understanding_2013}, an ability to couple and decouple qubits, and low hardware and operational complexity. In the platform of circuit QED~\cite{blais_cavity_2004}, two-qubit gates with errors below the 1\% threshold for surface code-based quantum error correction~\cite{fowler_surface_2012} have been demonstrated using multiple schemes. Approaches utilizing qubits and/or couplers with flux-tunable transition frequencies have shown high tunability~\cite{collodo_implementation_2020,xu_high-fidelity_2020,sung_realization_2020} and cancellation of idle coupling between qubits~\cite{mundada_suppression_2019,sung_realization_2020}, and recently schemes using tunable qubits, couplers, or both have reported fidelities above $99.8\%$~\cite{negirneac_high-fidelity_2020,stehlik_tunable_2021,sung_realization_2020}. However, these approaches require additional control lines and circuitry, and introduce decoherence channels.

A scheme with reduced hardware complexity that allows for high fidelity, tunable coupling and decoupling of qubits is of interest for scaling up quantum computers. Charge-activated two-qubit gates~\cite{paraoanu_microwave-induced_2006, chow_microwave-activated_2013, puri_high-fidelity_2016, paik_experimental_2016, rigetti_fully_2010, long_universal_2021, krinner_demonstration_2020} offer hardware simplicity, as they do not require additional control lines, and they are compatible with fixed-frequency qubits, which exhibit higher coherence times. Of the charge-activated gates, the cross resonance (CR) gate~\cite{rigetti_fully_2010} has reported the highest fidelity of $99.7\%$~\cite{kandala_demonstration_2020}. However, schemes relying on fixed-frequency, fixed-coupling architectures often have residual idle $ZZ$ coupling between qubits~\cite{dicarlo_demonstration_2009}, where $Z$ is the single-qubit Pauli $\sigma_z$-operator, due to interactions between noncomputational transitions of the superconducting qubits. This residual coupling can cause correlated errors, dephasing, and spectator errors ~\cite{gambetta_characterization_2012, mckay_three-qubit_2019, morvan_qutrit_2020, krinner_benchmarking_2020}. Mitigating the impact of the idle $ZZ$ coupling without a flux-tunable coupler has been done using dynamical decoupling techniques~\cite{sheldon_procedure_2016, sundaresan_reducing_2020,long_universal_2021} and using opposite-anharmonicity qubits~\cite{ku_suppression_2020}, however these approaches add overhead in circuit depth and in hardware complexity, respectively. Recently, cancellation of the idle $ZZ$ coupling between fixed-frequency, capacitively-coupled fluxonium qubits was also reported~\cite{xiong_arbitrary_2021}.

\begin{figure}[!t]
    \centering
    \includegraphics[width=8cm]{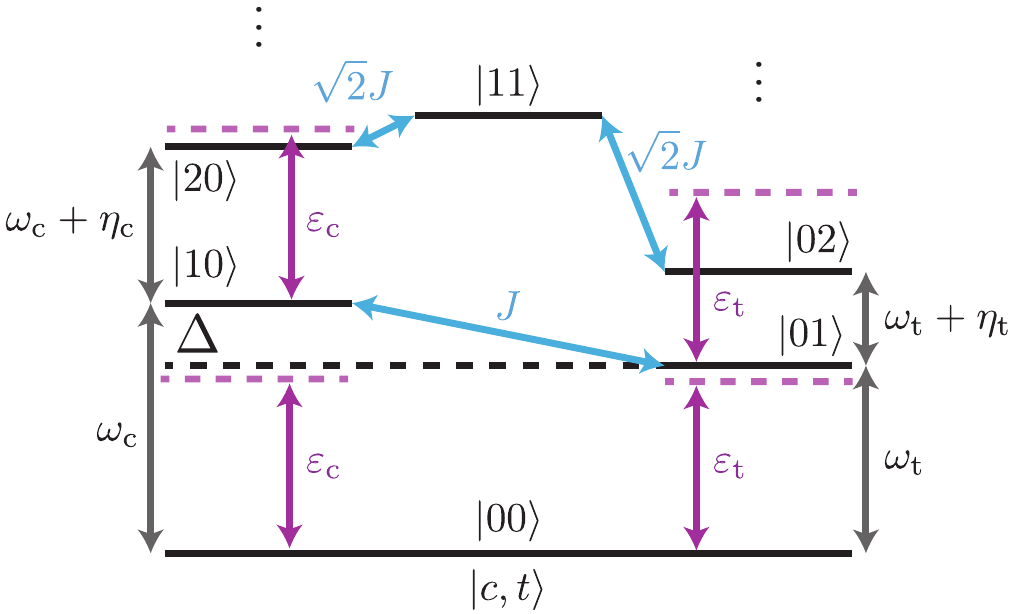}
    \caption{\textbf{Drive Scheme for the Stark-induced ZZ interaction}. Transmons are detuned by $\Delta = \omega_{\mathrm{c}} - \omega_{\mathrm{t}}$ indicated by the black dashed line, and coupled via exchange coupling $J$. They are driven simultaneously with amplitudes $\varepsilon_\mathrm{c}$, $\varepsilon_\mathrm{t}$ between frequencies  $\omega_\mathrm{t}$ and $\omega_\mathrm{c} + \eta_c$, indicated with purple dashed lines. The simultaneous driving introduces conditional Stark shifts, i.e., a $ZZ$ interaction.}
    \label{fig:drive_diagram}
\end{figure}

In this Letter, we demonstrate a tunable, charge-activated $ZZ$ interaction between two fixed-frequency transmon qubits~\cite{koch_charge-insensitive_2007} with fixed dispersive coupling, enabling full cancellation of idle $ZZ$ coupling between the qubits, and further realize a CZ gate with a fidelity of $99.43(1)\%$. The gate is realized by driving the transmons simultaneously, at a frequency between the $\ket{0}\rightarrow\ket{1}$ and $\ket{1}\rightarrow\ket{2}$ transitions, causing state-dependent Stark shifts due to the state-dependent frequency detunings from the drive frequency (see Fig.~\ref{fig:drive_diagram}).  The interaction can be activated with a range of drive frequencies, allowing one to avoid driving unwanted transitions in a crowded frequency spectrum of a quantum processor~\cite{hertzberg_laser-annealing_2020,zhang_high-fidelity_2020}. Further, \textit{in situ} control over the dispersive coupling between fixed-frequency, fixed-coupling transmons enables controlled phase gates with arbitrary phase angles $\mathrm{CZ}\left(\phi\right)$, which are useful for noisy intermediate-scale quantum (NISQ) algorithms~\cite{preskill_quantum_2018}, e.g., variational quantum algorithms~\cite{lacroix_improving_2020, foxen_demonstrating_2020}.

Consider two coupled, simultaneously driven transmons, as illustrated in Figure~\ref{fig:drive_diagram}. The Hamiltonian of the two transmons, in the frame of the drive at frequency $\omega_{\mathrm{d}}$ and making the Duffing approximation of the transmon~\cite{koch_charge-insensitive_2007}, is given by 
\begin{equation}
    H_{\mathrm{qb}} =  \sum_{i=\mathrm{c,t}}\left(\omega_{i} - \omega_{\mathrm{d}}\right)a_i^\dagger a_i + \frac{\eta_i}{2}a_i^\dagger a_i^\dagger a_i a_i,
\end{equation}
where for transmon $i$, $a_i$ is the bosonic annihilation operator, $\omega_{i}$ is the transition frequency between $\ket{0}$ and $\ket{1}$, $\eta_i$ is the anharmonicity, and $\hbar=1$. Each drive term is given by $H_{\varepsilon_i} = \left(\varepsilon_i a_i +\varepsilon_i^* a_i^\dag\right)$, where $\varepsilon_i$ is the complex drive amplitude, and the coupling term with coupling strength $J$ is $H_J = J\left(a_\mathrm{c}^\dag a_\mathrm{t} + a_\mathrm{c} a_\mathrm{t}^\dag\right)$, where we denote the higher frequency transmon $Q_\mathrm{c}$, and the lower frequency transmon as $Q_\mathrm{t}$. The total system Hamiltonian is then
$H = H_{\mathrm{qb}} + H_J + H_{\varepsilon_\mathrm{c}} + H_{\varepsilon_\mathrm{t}}$. 

The Stark-induced $ZZ$ interaction can be understood through the lens of the CR effect: when driving the control qubit with amplitude  $\varepsilon_\mathrm{c}$ at the target qubit frequency $\omega_{\mathrm{t}}$, the target qubit experiences a drive amplitude  $\tilde{\varepsilon}_n$ that depends on the control qubit state $\ket{n}$~\cite{tripathi_operation_2019}. In the qubit subspace, the CR drive then realizes an entangling $ZX$ interaction with interaction rate  $\mu = \left(\tilde{\varepsilon}_0 - \tilde{\varepsilon}_1\right)/2$. In the limit $\varepsilon_\mathrm{c}/\Delta_\mathrm{t} \ll 1$, detuning the drive frequency from the target qubit frequency $\Delta_\mathrm{t} = \omega_{\mathrm{t}} - \omega_{\mathrm{d}}$ results in a conditional off-resonant drive, or a conditional Stark shift of the target qubit frequency $\tilde{\delta}_n$, where
\begin{equation}
    \tilde{\delta}_n = \frac{\tilde{\varepsilon}_n^2}{\Delta_\mathrm{t}}.
\end{equation}
The drive-induced $ZZ$ interaction $\zeta$ then is given by $\zeta = \tilde{\delta_0} - \tilde{\delta_1}$, which can be expressed in terms of $\mu$ as
\begin{equation}
    \zeta=2\mu\left(\tilde{\varepsilon}_0 + \tilde{\varepsilon}_1\right)/\Delta_\mathrm{t}.
\end{equation}

In the off-resonant limit, the conditional Stark shift is much smaller than the CR rate $\zeta \ll \mu$. However, by applying a drive tone simultaneously to the target qubit at amplitude $\varepsilon_\mathrm{t}$, the drive amplitude on the target becomes $\tilde{\varepsilon}_n + \varepsilon_\mathrm{t}$, and therefore $\zeta$ is enhanced by the (unconditional) increase in drive amplitude. By replacing $\tilde{\varepsilon}_n \rightarrow \tilde{\varepsilon}_n + \varepsilon_\mathrm{t}$ above, $\zeta$ then scales linearly to first order with $\varepsilon_\mathrm{t}$. 

\begin{align}
    \zeta = \frac{2\mu}{\Delta_\mathrm{t}}\left(\tilde{\varepsilon}_0 + \tilde{\varepsilon}_1  + 2\varepsilon_\mathrm{t}\right) + O(|\varepsilon_\mathrm{t}|^2)
\end{align}
Thus, by driving both transmons simultaneously, it is possible to realize a $ZZ$ interaction with comparable interaction rates as the CR effect when using fixed-frequency, fixed-coupling transmons that are in a similar detuning $\Delta$ as those designed for the CR gate. Note that in the above description, the qubits are interchangeable, and this description can be generalized to systems that are not transmons~\cite{xiong_arbitrary_2021}. A formal derivation of the effect using perturbation theory is given in the supplementary material.

Motivated by the model described above, we experimentally investigate how the Stark-induced $ZZ$ interaction depends on the field amplitudes on each qubit $\varepsilon_{\mathrm{c}}$, $\varepsilon_{\mathrm{t}}$ and the frequency of the drive field $\omega_{\mathrm{d}}$.  We use Ramsey interferometry~\cite{ramsey_molecular_1950} and extract the frequency shift of $Q_{\mathrm{t}}$ conditioned on the state $Q_{\mathrm{c}}$ being $\ket{0}$ or $\ket{1}$ when applying the off-resonant drive. We find agreement between our measurements and numerical simulations when microwave crosstalk is included.

\begin{figure}[!th]
    \centering
    \begin{tikzpicture}
        \node[anchor=north west,inner sep=0] (image) at (0.0,-0.1cm) {\includegraphics[width=7.0cm]{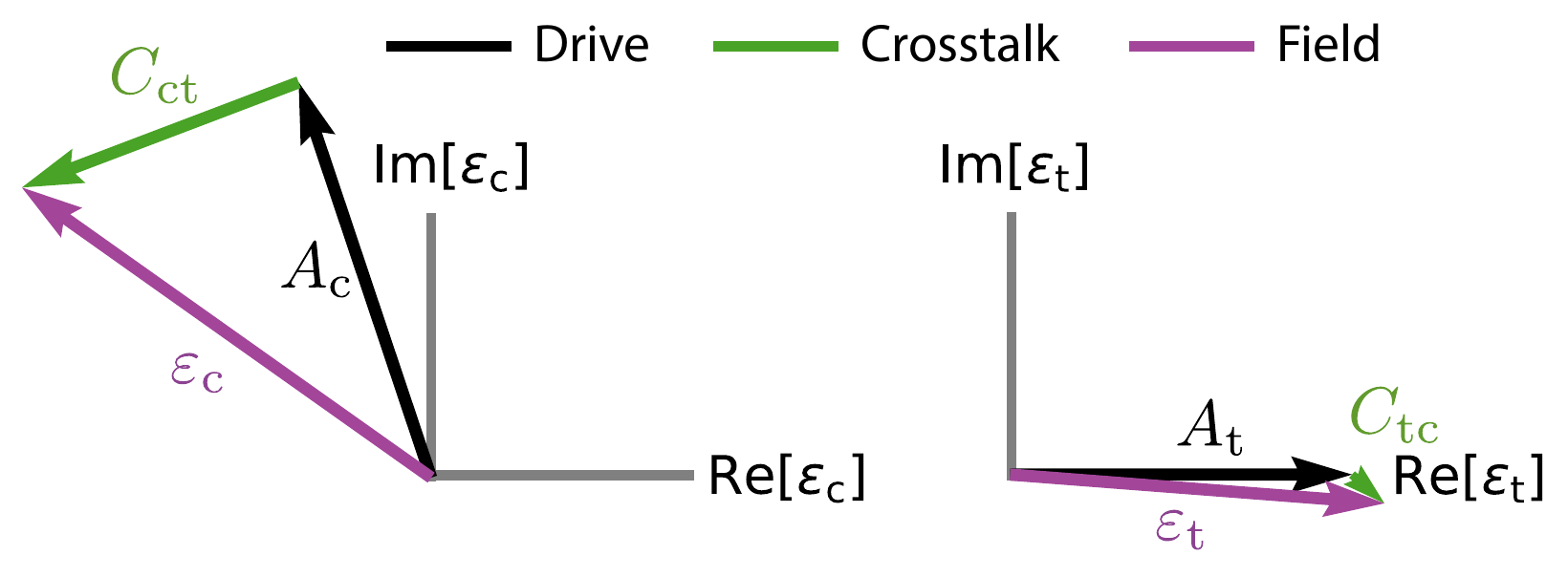}};
        \node[] at (0.0,-0.1) {\textbf{(a)}}; 
        \node[anchor=north west,inner sep=0] (image) at (0.0, -3.2cm)
        {\includegraphics[width=7cm]{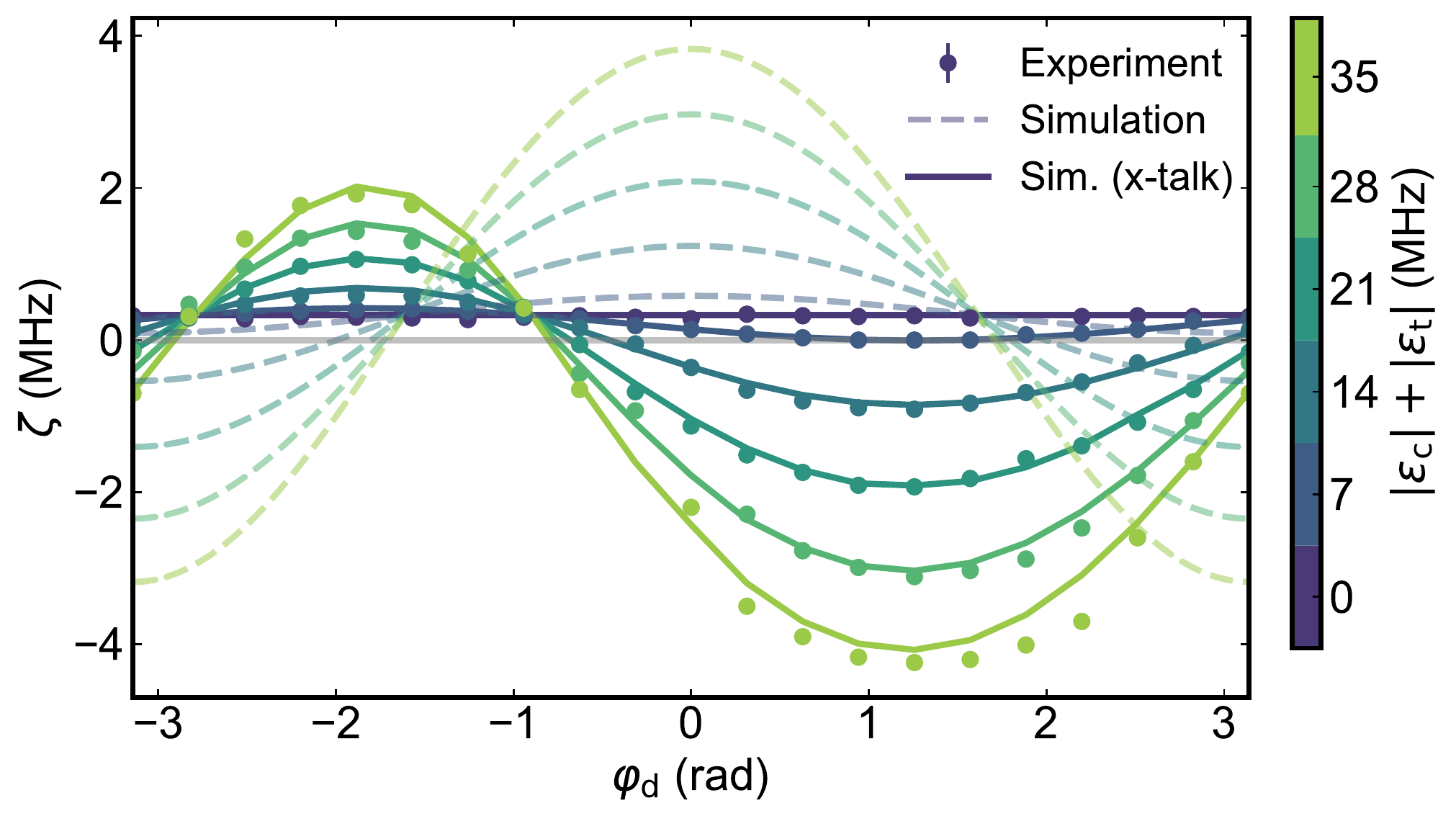}};
        \node[] at (0.0, -3.2cm) {\textbf{(b)}}; 
        \node[anchor=north west,inner sep=0] (image) at (0.0,-7.2cm) {\includegraphics[width=7cm]{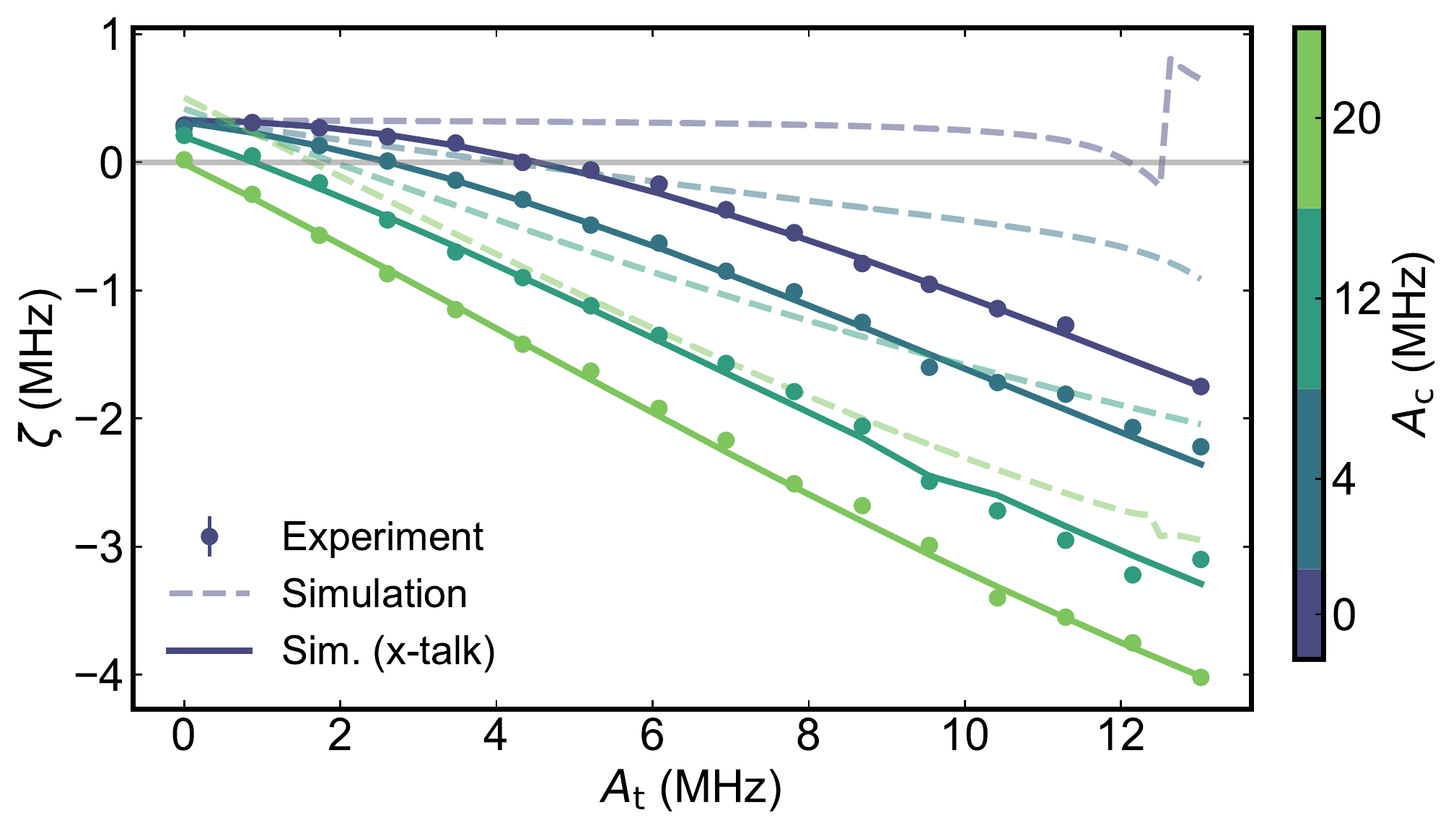}};
        \node[] at (0.0,-7.2cm) {\textbf{(c)}}; 
    \end{tikzpicture}
    \caption{\textbf{$ZZ$ as a function of relative drive phase, crosstalk} \textbf{(a)}  The fields $\varepsilon_{\mathrm{c}}$, $\varepsilon_{\mathrm{t}}$ incident on the control and target qubits respectively are complex combinations of the drive line amplitudes $A_{\mathrm{c}}$, $A_{\mathrm{t}}$, mixed via microwave crosstalk matrix $C$. \textbf{(b)} $ZZ$ versus relative drive phase $\varphi_\mathrm{d}$, for several overall drive amplitudes $|A|=|A_{\mathrm{c}}|=|A_{\mathrm{t}}|$. Fits to numerical models map the drive amplitudes to field amplitudes $\varepsilon_{\mathrm{c}}$, $\varepsilon_{\mathrm{t}}$, indicated in the colorbar. Microwave crosstalk results in an asymmetry in maximal magnitude of $ZZ$ between positive and negative signs, and a shift in the phase corresponding to the maximum and minimum $ZZ$ rates. Data are shown in dots, with error bars indicating Ramsey frequency fit uncertainty. Simulation with crosstalk is shown with solid lines, and crosstalk-free simulation is shown in dashed lines. \textbf{(c)} $ZZ$ versus target drive amplitude $A_{\mathrm{t}}$ for several control drive amplitudes $A_{\mathrm{c}}$, indicated in the colorbar. Microwave crosstalk results in larger $ZZ$ than crosstalk-free driving (dashed lines), and linear behavior with respect to target drive amplitude is observed when also driving the control transmon.}
    \label{fig:zz_xtalk}
\end{figure}

The experiments measuring $ZZ$ as a function of drive parameters, shown in Figure~\ref{fig:zz_xtalk}, were performed on a pair of fixed-frequency transmons with parameters $\omega_{\mathrm{c}}/2\pi=5.845\unit{GHz}$ ($\omega_{\mathrm{t}}/2\pi=5.690\unit{GHz}$), $\eta_{\mathrm{c}}/2\pi=-244.1\unit{MHz}$ ($\eta_{\mathrm{t}}/2\pi=-247.1\unit{MHz}$) with static $ZZ$ coupling $\zeta_0/2\pi=307\unit{kHz}$, corresponding to an inferred exchange coupling strength $J = 3.45\unit{MHz}$. The relaxation and coherence times of $Q_{\mathrm{c}}$ ($Q_{\mathrm{t}}$) are $T_1=80\unit{\mu s}$, ($T_1=100\unit{\mu s}$) and $T_2^{\mathrm{echo}}=150\unit{\mu s}$, ($T_2^{\mathrm{echo}}=180\unit{\mu s}$).

In the presence of microwave crosstalk, $\varepsilon_{\mathrm{c}}$ and $\varepsilon_{\mathrm{t}}$ are given by complex linear combinations of CZ pulse amplitudes $A_{\mathrm{c}}$ and $A_{\mathrm{t}}$ applied to the control and target transmon drive with relative phase $\varphi_{\mathrm{d}}$, as shown in Figure~\ref{fig:zz_xtalk} (a), according to
\begin{equation}
    \begin{pmatrix}
    \varepsilon_{\mathrm{c}} \\
    \varepsilon_{\mathrm{t}}
    \end{pmatrix} = 
    \begin{pmatrix}
    e^{i\theta_{\mathrm{c}}} & C_{\mathrm{ct}}e^{i\varphi_{\mathrm{ct}}} \\
    C_{\mathrm{tc}}e^{i\varphi_{\mathrm{tc}}} & 1
    \end{pmatrix} \begin{pmatrix}
    A_{\mathrm{c}} \\
    A_{\mathrm{t}} e^{-i\varphi_\mathrm{d}}
    \end{pmatrix},
\end{equation}
where $C_{\mathrm{ct}}$ ($\varphi_{\mathrm{ct}}$) denotes the crosstalk amplitude (phase) mapping $A_{\mathrm{t}}$ to $\varepsilon_{\mathrm{c}}$. The phase $\theta_{\mathrm{c}}$ results from differences in electrical delay between the drive lines. For these experiments, we set the drive detuning $\Delta_{\mathrm{t}} = 40\unit{MHz}$ and measured the $ZZ$ interaction as a function of drive phase $\varphi_\mathrm{d}$ and global drive amplitude $A$, where $A_{\mathrm{c}} = A_{\mathrm{t}} = A$, as shown in Fig.~\ref{fig:zz_xtalk} (b). The experimental data deviates from the crosstalk-free simulations, which diagonalize the full system Hamiltonian (see supplementary material), however when fitting the data to a model including crosstalk, the data agrees well with simulation. We further note that in the absence of crosstalk, the absolute interaction rates are symmetric about $\zeta_0$ between in-phase and out-of-phase driving. In addition to sweeping relative phase, we varied $A_{\mathrm{t}}$ and $A_{\mathrm{c}}$ independently with drive phase fixed to the phase where we measured maximal $|\zeta|$, $\varphi_{\mathrm{d}}=1.31\unit{rad}$. These results are shown in Fig.~\ref{fig:zz_xtalk} (c), where the linear dependence of $\zeta$ on the drive amplitude is observed for non-zero $A_{\mathrm{c}}$ amplitudes, as was predicted in the model presented above.

Together, these experiments demonstrate wide tunability of the $ZZ$ interaction, from $|\zeta|$ one order of magnitude larger than $\zeta_0$, as well as freedom to cancel $\zeta$ by adjusting the relative phase between the drives. This tunability is available even in the presence of crosstalk, and the scaling of $\zeta$ with respect to drive parameters agrees well with numerical simulation. 

\begin{figure}[!tbp]
    \centering
    \begin{tikzpicture}
        \node[anchor=north west,inner sep=0] (image) at (1.0,0.0cm) {\includegraphics[width=6cm]{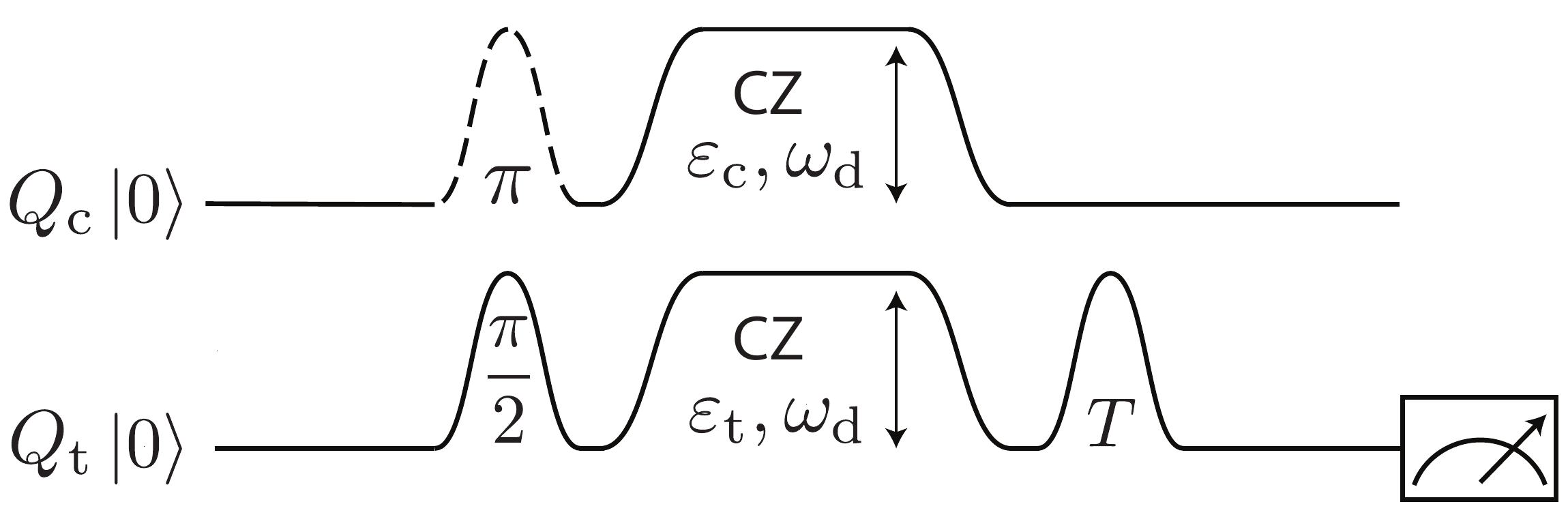}};
        \node[anchor=north west,inner sep=0] (image) at (0.0,-2.2cm) {\includegraphics[width=8cm]{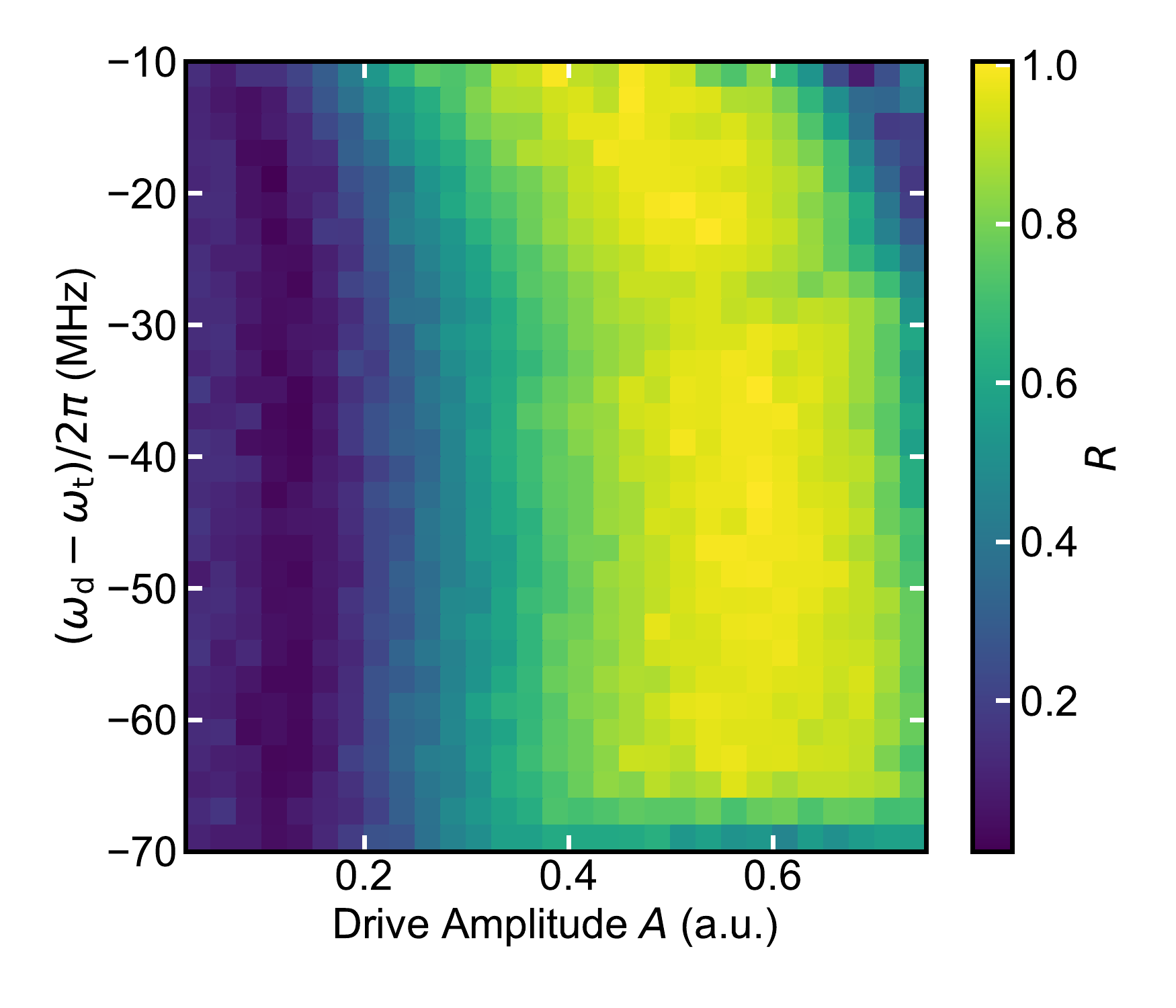}};
        \node[] at (0.0,0.0) {\textbf{(a)}}; 
        \node[] at (0.0,-2.2cm) {\textbf{(b)}}; 
    
    \end{tikzpicture}
    \caption{\textbf{CZ Gate Calibration} \textbf{(a)} Pulse sequence for calibrating the amplitude and frequency of the CZ pulse. The pulse entangles the qubits when the target qubit is prepared along the equator. Tomographic pulses $T$ are applied to measure the target qubit Bloch vector $r_i$ for each control qubit state $\ket{i}\in \{0,1\}$ and extract entanglement measure $R$, similar to the measure used in~\cite{sheldon_characterizing_2016}. The global pulse amplitude and frequency are calibrated by selecting parameters that maximize $R$. \textbf{(b)}  A sweep of $R$ as a function of the CZ gate amplitude $A$ and drive detuning from the target $(\omega_{\mathrm{d}} - \omega_{\mathrm{t}})$. There is a band of frequencies available where $R$ is maximal to realize the CZ gate. There is also a pulse amplitude around $A=0.1$ where entanglement is minimal, corresponding to $ZZ$ cancellation.}
    \label{fig:R_cal2D}
\end{figure}

We next apply the strong $ZZ$ interactions to calibrating a CZ gate, which is realized with Hamiltonian terms $IZ$, $ZI$, and $ZZ$: $\mathrm{CZ} = \exp \left(-\frac{i}{2}\frac{\pi}{2} \left(-ZI - IZ + ZZ\right)\right)$. We first calibrate the entangling term $ZZ$, and then correct local phase errors on the each qubit using virtual $Z$ gates~\cite{mckay_efficient_2017}. To calibrate the $ZZ$ term, we prepare the target qubit along the equator in the Bloch sphere, apply the CZ pulse, and measure the target qubit Bloch vector $r_0$ ($r_1$) when the control qubit is in the $\ket{0}$ ($\ket{1}$) state (see Fig.~\ref{fig:R_cal2D} (a)). To maximize entanglement, we maximize the quantity $R$, similar to what was introduced in ~\cite{sheldon_procedure_2016}
\begin{equation}
    R = \frac{1}{2}||r_0 - r_1||^2,
\end{equation}
which measures the normalized vector distance between target Bloch vectors conditioned on the control qubit state. For the drive pulse, we use a cosine ramp with a flat top. The pulse parameters to be calibrated are: pulse time $\tau_p$, drive frequency $\omega_{\mathrm{d}}$, the drive amplitude applied to each qubit $A_{\mathrm{c}}$ and $A_{\mathrm{t}}$, relative phase between each drive $\varphi_{\mathrm{d}}$. We chose the duration of the flat top of the pulse to be $40\%$ of the total pulse time. Pulse time selection is informed by experiments sweeping drive amplitudes and relative phase, like that shown in Figure~\ref{fig:zz_xtalk}. 
We calibrate the gate on a separate qubit pair than was used for the preceding experiments, with parameters $\omega_{\mathrm{c}}/2\pi=5.4696\unit{GHz}$ ($\omega_{\mathrm{t}}/2\pi=5.315\unit{GHz}$), $\eta_{\mathrm{c}}/2\pi=-270.5\unit{MHz}$ ($\eta_{\mathrm{t}}/2\pi=-273.0\unit{MHz}$) with static $ZZ$ coupling $\zeta_0/2\pi=170\unit{kHz}$, corresponding to an inferred exchange coupling strength $J = 2.79\unit{MHz}$. The relaxation and coherence times of each transmon are $T_1=65(5)\unit{\mu s}$ ($T_1=58(9)\unit{\mu s}$) and $T_2^{\mathrm{echo}}=86(6)\unit{\mu s}$, ($T_2^{\mathrm{echo}}=77(8)\unit{\mu s}$). For the gate discussed, we set $\tau_p=201\unit{ns}$. We select pulse amplitudes $A_{\mathrm{c}}$, $A_{\mathrm{t}}$ and relative phase $\varphi_{\mathrm{d}}$ that maximize $|\zeta|$.

To calibrate the drive frequency and amplitude, we measure $R$ vs global pulse amplitude $A$ and drive frequency $\omega_{\mathrm{d}}$, as shown in Figure~\ref{fig:R_cal2D}. One observes a bandwidth of $40\unit{MHz}$ where maximal $R$ is achievable. In this region, the amplitude is insensitive to the frequency, consistent with the off-resonant nature of the interaction. We then calibrate local phase corrections by measuring the individual qubit Pauli $Z$ error using Ramsey type experiments to map the pulse to a CZ gate. We further note that there is a region where $R$ nears zero, corresponding to $ZZ$ interaction cancellation, as shown in Fig.~\ref{fig:zz_xtalk} (b).

\begin{figure}[!tbp]
    \centering
    \begin{tikzpicture}
        
        \node[anchor=north west,inner sep=0] (image) at (0.0,0.0cm) {\includegraphics[width=8.25cm]{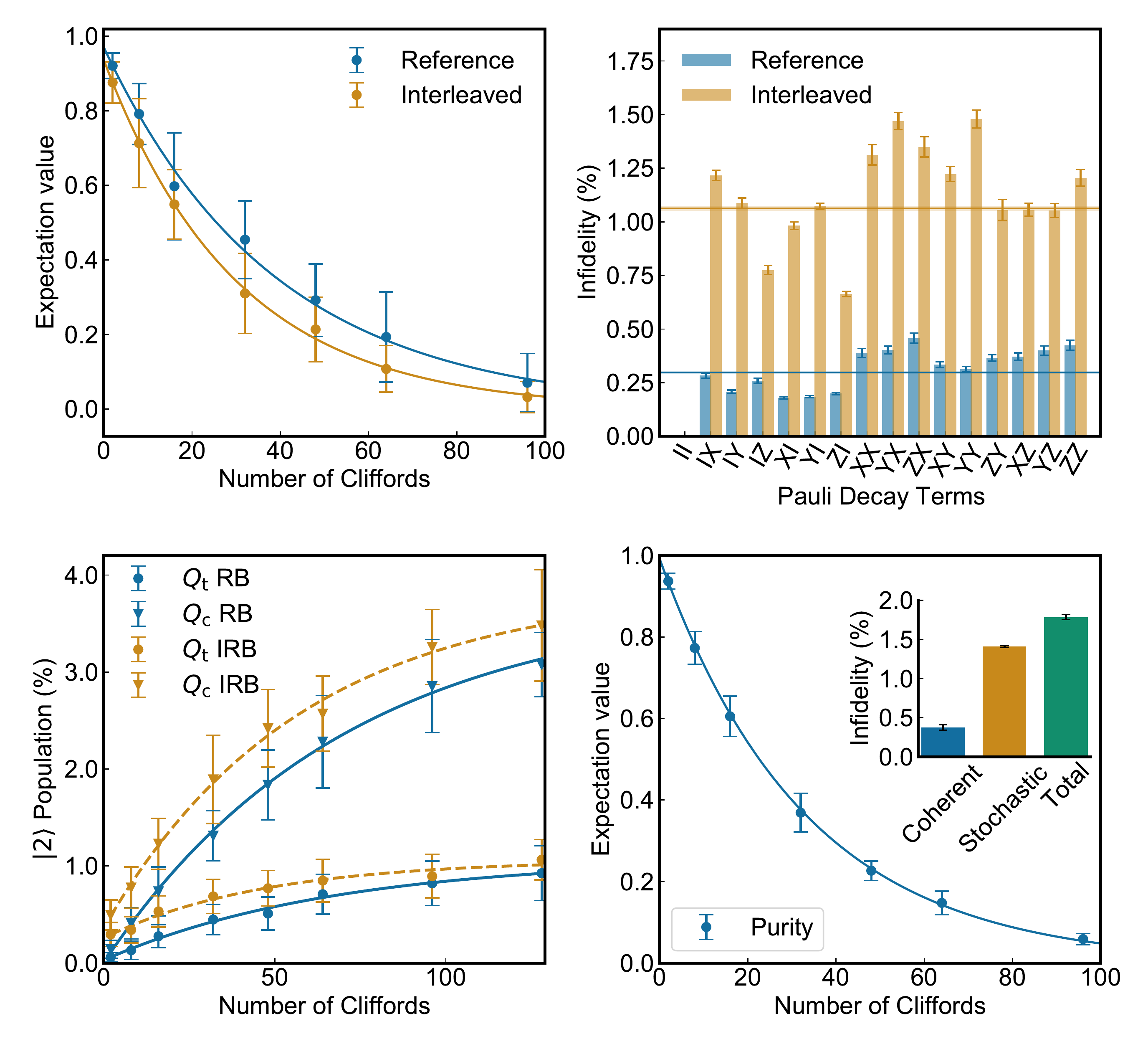}};
        \node[] at (0.0,0.0) {\textbf{(a)}}; 
        \node[] at (0.0, -3.75cm) {\textbf{(c)}}; 
        \node[] at (4.45cm,0.0) {\textbf{(b)}}; 
        \node[] at (4.45cm, -3.75cm) {\textbf{(d)}}; 
        
    \end{tikzpicture}
    \caption{\textbf{Benchmarking Results} \textbf{(a)} Interleaved RB. Exponential decay of Clifford sequence fidelity (y-axis) for different Clifford sequence lengths with (gold) and without (blue) the interleaved CZ gate. From reference and interleaved decay parameters $p_{RB}=0.9744(9)$ and $p_{IRB}=0.9672(7)$, respectively, we extract a CZ gate fidelity of $99.44(9)\%$.  \textbf{(b)} Cycle Benchmarking results. The error rate $e_i$ is obtained for the prepared Pauli eigenstate $i$ from the exponential decay of sequence fidelity $p_i$ with increasing sequence length. A larger error rate for a given Pauli eigenstate indicates errors in the cycle that do not commute with the compiled Pauli term. The process infidelity is obtained by averaging across all Pauli terms. By performing CB for the CZ gate (gold) and the identity cycle (blue), we extract a gate fidelity of $99.43(1)\%$.  \textbf{(c)} Leakage Randomized Benchmarking. By monitoring the $\ket{2}$ state of each transmon when running an IRB experiment, the reference and interleaved $\ket{2}$ state population data are each fit to an exponential model to extract the leakage-per-gate for each transmon. \textbf{(d)} Purity Benchmarking. This experiment distinguishes coherent from stochastic errors by measuring the decay of the purity (main plot) of the two-qubit density matrix by performing state tomography for each random Clifford RB sequence, and comparing the purity decay to the RB decay. Inset: breakdown of Clifford process infidelity between coherent and stochastic contributions. We find stochastic error to be the dominant source of error.}
    
    \label{fig:benchmarking}
\end{figure}
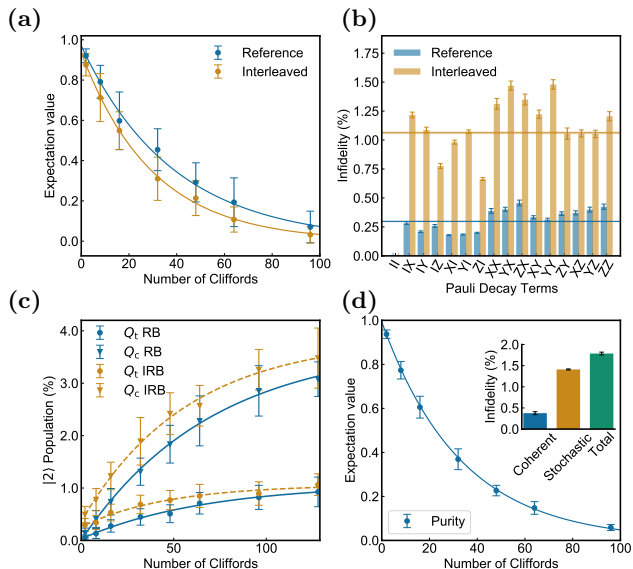

To quantify the fidelity of the gate and sources of error, we perform a suite of randomized benchmarking~\cite{emerson_scalable_2005,knill_randomized_2008} experiments. We first perform Interleaved Randomized Benchmarking (IRB)~\cite{magesan_efficient_2012}, shown in Figure~\ref{fig:benchmarking} (a). The IRB protocol interleaves a gate of interest between a sequence of randomly chosen Clifford gates, a method called Clifford twirling, which randomizes gate errors to a depolarizing channel. The sequence fidelity is measured with increasing sequence length $m$ and the decay parameter $p$ is extracted by fitting to the exponential model $P(m) = A\cdot p^m$~\cite{harper_statistical_2019}. Because the experiments estimate $F_{CZ}$ based on exponential decays, IRB is insensitive to state preparation and measurement (SPAM) errors. The decay parameter for the interleaved experiment $p_{IRB}=0.9672(7)$ and the reference experiment with no interleaved gate $p_{RB}=0.9744(9)$ give an estimate of the CZ gate fidelity $F_{CZ} = \frac{d-1}{d}\left(1-p_{IRB}/p_{RB}\right)=99.44(9)\%$, where $d=2^n$ for $n$ qubits (for our experiments $d=4$). The upper- and lower-bounds on gate fidelity estimates from IRB have been shown to span orders of magnitude~\cite{carignan-dugas_bounding_2019}. In general, when the decay of the reference experiment fidelity is comparable to the interleaved experiment fidelity, the IRB estimate  uncertainty increases. For our experiments, the upper- and lower-bounds on $F_{CZ}$ from IRB are between $91.9(2)\%$ and $99.96(1)\%$, spanning roughly 2 orders of magnitude in gate error.

To reduce gate fidelity estimate uncertainty, we run the cycle benchmarking (CB) protocol~\cite{erhard_characterizing_2019}, shown in Figure~\ref{fig:benchmarking} (b). The CB protocol is similar to IRB, in that the gate, or cycle, of interest is interleaved between randomly chosen gates. Instead of Clifford gates, in CB the cycle is twirled with multi-qubit Pauli gates, which are simply tensor products of single-qubit Pauli gates. Pauli twirling maps gate errors into stochastic Pauli errors, which are measured by preparing each eigenstate of the two-qubit Pauli basis, e.g. $XX$ or $YZ$, on the system, and fitting the sequence fidelity to an exponential decay parameter as in RB (e.g., $p_{XX}$, $p_{YZ}$) as a function of CB sequence length. A larger error rate $e_i=(1-p_i)(1-1/d^2)$ for a given Pauli eigenstate $i$ indicates errors in the cycle that do not commute with that Pauli operator. We performed CB with cycle lengths of $m\in \{2,16,32\}$ for the CZ cycle and for the identity cycle (as a reference for the fidelity of the Pauli twirling gates) and extracted the error rate of each Pauli term, which is plotted in Figure~\ref{fig:benchmarking} (b). Averaging over all Pauli preparations, we extract average cycle process fidelities of $p_{CZ}=0.98937(8)$ and $p_I=0.99702(3)$. We then estimate the CZ gate fidelity as in the IRB protocol, to be $F_{CZ} = \frac{d-1}{d}\left(1-p_{CZ}/p_I\right)=99.43(1)\%$, with a worst-case fidelity bound of $97.52(2)\%$ and a best-case fidelity bound of $99.764(5)\%$. Note that the worst-case and best-case fidelity bounds from CB are narrower than that of IRB, due to the higher fidelity of the Pauli twirling operation over Clifford twirling.

To understand how to reduce the CZ gate error, it is important to distinguish the different error sources. Different error types include coherent errors, such as mis-calibration, stochastic errors, such as dephasing or relaxation errors, and leakage errors, involving population transfer to non-computational states of the system. We measure the leakage-per-gate using leakage randomized benchmarking (LRB)~\cite{wood_quantification_2018, chen_measuring_2016}, realized by extracting $\ket{2}$-state outcomes from IRB experiments. By fitting the qubit state populations data shown in Figure~\ref{fig:benchmarking} (c) to a rate-equation model~\cite{chen_measuring_2016,wood_quantification_2018} for both reference RB and interleaved RB experiments, we resolve the leakage-per-gate for each transmon to be $0.014\%$ and $0.007\%$ for $Q_\mathrm{c}$ and $Q_\mathrm{t}$ respectively, indicating leakage is not a dominant source of errors for this gate.

To distinguish coherent and stochastic error sources, we perform purity benchmarking~\cite{wallman_estimating_2015}, using the Extended Randomized Benchmarking (XRB) protocol~\cite{beale_true-q_2020}, which measures the decay of the purity of the two-qubit density matrix by performing state tomography after Clifford RB sequences. Figure~\ref{fig:benchmarking} (d) shows the purity decay curve, and the inset shows the breakdown of the Clifford process infidelity  $e_F=\left(1-p_{RB}\right)\left(1-1/d^2\right)= 1.78(3)\cdot 10^{-2}$ between coherent ($e_U =0.37(3)\cdot 10^{-2}$) and stochastic ($e_S=1.41(1)\cdot 10^{-2}$) error types. The dominant source of error in the gate is stochastic error. We estimate the decoherence-limited Clifford process infidelity $e_{\mathrm{decoh.}}$ using the $T_1$ and $T_2^{\mathrm{echo}}$  values and the average Clifford gate length of $389\unit{ns}$, which gives $e_{\mathrm{decoh.}}=0.76\cdot 10^{-2}$. The larger observed $e_S$ than the $e_{\mathrm{decoh.}}$ suggests that other forms of stochastic error are present beyond those introduced by relaxation and dephasing of the qubit transition levels. This could be due to decoherence channels of higher transmon levels participating in the interaction via state hybridization between computational and noncomputational levels during the drive~\cite{xiong_arbitrary_2021}. We have further evidence of reduced coherence when the drive is on in the supplementary material. To summarize, we measure the CZ gate fidelity to be $99.43(1)\%$, with low leakage and the dominant source of remaining errors being stochastic error. The discrepancy between stochastic error and predicted error from decoherence suggests additional sources of stochastic error are present during the gate.

We have demonstrated a tunable $ZZ$ interaction between fixed frequency, fixed coupling transmons using off-resonant, simultaneous charge drives. This tunable $ZZ$ coupling enables both cancellation and enhancement, of the static $ZZ$ interaction. In this work, we implemented a high fidelity CZ gate that is resilient against drive crosstalk and static $ZZ$ interactions during the gate. We expect with this interaction one can leverage higher exchange coupling $J$ between qubits to further reduce CZ gate times, and  multi-path couplers that suppress static $ZZ$~\cite{kandala_demonstration_2020} can be combined with this drive to eliminate the unwanted $ZZ$ during idling~\cite{mundada_suppression_2019} or during operation of other gates. Further, the off-resonant character of the interaction provides drive frequency flexibility, reducing frequency crowding constraints with scaling up fixed frequency, fixed coupling quantum processors. While we utilize this charge-activated tunable $ZZ$ interaction to implement a CZ gate, it can also be applied to simulation of exotic quantum many-body physics, including the extended Bose-Hubbard model~\cite{kounalakis_tuneable_2018,roushan_spectroscopic_2017, ye_propagation_2019,jin_photon_2013} and nonreciprocal interacting photonic systems~\cite{roushan_chiral_2017,guan_synthetic_2020}, as well as realizing native quantum stabilizer measurements~\cite{andersen_repeated_2020}, without the need for additional flux-based tunable components.

We thank Larry Chen, Trevor Chistolini, William Livingston, Dr.\ Long Nguyen, and Dr.\ Jean-Loup Ville for valuable discussions. This material is based upon work supported in part by the U.S. Army Research Laboratory and the U.S. Army Research Office under contract/grant number W911NF-17-S-0008 and the National Defense Science \& Engineering Graduate (NDSEG) Fellowship.

\clearpage
\bibliographystyle{apsrev4-1}
\bibliography{cz_refs}

\end{document}


\title{Supplementary Material for ``Hardware-Efficient Microwave-Activated Tunable Coupling Between Superconducting Qubits"}

\author{Bradley K. Mitchell}
\affiliation{Quantum Nanoelectronics Laboratory, University of California, Berkeley, Berkeley CA 94720}
\affiliation{Computational Research Division, Lawrence Berkeley National Laboratory, Berkeley CA 94720}
\author{Ravi K. Naik}
\affiliation{Quantum Nanoelectronics Laboratory, University of California, Berkeley, Berkeley CA 94720}
\affiliation{Computational Research Division, Lawrence Berkeley National Laboratory, Berkeley CA 94720}
\author{Alexis Morvan}
\affiliation{Quantum Nanoelectronics Laboratory, University of California, Berkeley, Berkeley CA 94720}
\affiliation{Computational Research Division, Lawrence Berkeley National Laboratory, Berkeley CA 94720}
\author{Akel Hashim}
\affiliation{Quantum Nanoelectronics Laboratory, University of California, Berkeley, Berkeley CA 94720}
\affiliation{Computational Research Division, Lawrence Berkeley National Laboratory, Berkeley CA 94720}
\author{John Mark Kreikebaum}
\affiliation{Quantum Nanoelectronics Laboratory, University of California, Berkeley, Berkeley CA 94720}
\affiliation{Materials Science Division, Lawrence Berkeley National Laboratory, Berkeley CA 94720}
\author{Brian Marinelli}
\affiliation{Quantum Nanoelectronics Laboratory, University of California, Berkeley, Berkeley CA 94720}
\affiliation{Computational Research Division, Lawrence Berkeley National Laboratory, Berkeley CA 94720}
\author{Wim Lavrijsen}
\affiliation{Computational Research Division, Lawrence Berkeley National Laboratory, Berkeley CA 94720}
\author{Kasra Nowrouzi}
\affiliation{Quantum Nanoelectronics Laboratory, University of California, Berkeley, Berkeley CA 94720}
\affiliation{Computational Research Division, Lawrence Berkeley National Laboratory, Berkeley CA 94720}
\author{David Santiago}
\affiliation{Quantum Nanoelectronics Laboratory, University of California, Berkeley, Berkeley CA 94720}
\affiliation{Computational Research Division, Lawrence Berkeley National Laboratory, Berkeley CA 94720}
\author{Irfan Siddiqi}
\affiliation{Quantum Nanoelectronics Laboratory, University of California, Berkeley, Berkeley CA 94720}
\affiliation{Computational Research Division, Lawrence Berkeley National Laboratory, Berkeley CA 94720}
\affiliation{Materials Science Division, Lawrence Berkeley National Laboratory, Berkeley CA 94720}

\maketitle

\section{Perturbation Theory for Conditional Stark Interaction}

The Hamiltonian of the system in the frame of the drive at frequency $\omega_{d}$ and approximating the transmons as Duffing oscillators is given in the main text and reproduced here,
\begin{equation}
    \label{eq:Hsupp}
    H=\sum_{i=\mathrm{c,t}}\left[\left(\omega_{i}-\omega_{\mathrm{d}}\right)a_{i}^{\dag}a_{i}+\frac{\eta_{i}}{2}a_{i}^{\dag}a_{i}^{\dag}a_{i}a_{i}+\varepsilon_{i}\left(e^{i\varphi_{i}}a_{i}+e^{-i\varphi_{i}}a_{i}^{\dag}\right)\right]+J\left(a_{\mathrm{c}}^{\dag}a_{\mathrm{t}}+a_{\mathrm{c}}a_{\mathrm{t}}^{\dag}\right),
\end{equation}
with $\hbar=1$ and the basis chosen such that $\varphi_{\mathrm{c}}=0$ and $\varphi_{\mathrm{d}}=\varphi_{\mathrm{t}}-\varphi_{\mathrm{c}}$. For convenience we define $\Delta_{i}=\omega_{i}-\omega_{\mathrm{d}}$, the detuning of transmon $i$ from the drive. The bare transmon Hamiltonians serve as the unperturbed system, $H_{0}$, and the perturbation, $V$ is composed of the single qubit drive terms and the coupling term. This is valid when $\varepsilon_{i},J \ll |\eta_{i}|,|\Delta_{i}|$. The $ZZ$ interaction rate between qubits is defined as 
\begin{equation}
    \zeta=E_{11}+E_{00}-\left(E_{01}+E_{10}\right)
\end{equation}
where $E_{ij}$ is the energy of the system with $Q_{\mathrm{c}}$ in state $i$ and $Q_{\mathrm{t}}$ in state $j$, denoted $\ket{ij}$. Using time independent perturbation theory we expand $\zeta=\zeta^{(0)}+\zeta^{(1)}+\cdots$ where $\zeta^{(n)}=E^{(n)}_{11}+E^{(n)}_{00}-(E^{(n)}_{01}+E^{(n)}_{10})$ and $E^{(n)}_{ij}$ is the $n$th order correction to the energy of the state $\ket{ij}$. In the unperturbed system the transmons are not interacting so $\zeta^{(0)}=0$. The perturbation $V$ maps states $\ket{ij}$ to states which are orthogonal to $\ket{ij}$ so there is no first order correction to the energies and $\zeta^{(1)}=0$ as well. 

At second order there are corrections to the energies from two sources. The coupling between the transmons produces a static $ZZ$ interaction and the drives on both transmons produce unconditional AC Stark shifts. The unconditional Stark shifts cancel out in the calculation of $\zeta^{(2)}$ which yields the familiar form of the static $ZZ$ interaction
\begin{equation}
    \zeta^{(2)}=2J^{2}\left(\frac{1}{\Delta-\eta_{\mathrm{t}}}-\frac{1}{\Delta+\eta_{\mathrm{c}}}\right),
\end{equation}
where $\Delta=\omega_{\mathrm{c}}-\omega_{\mathrm{t}}$ is the detuning between the control and target. 

At third order it becomes clear why simultaneously driving both transmons increases the $ZZ$ rate considerably relative to the case where only one transmon is driven. This can be seen by noting that the drive terms in the perturbation add or remove excitations from the system while the interaction term conserves the excitation number. In order to find a third order correction we need a process where a drive adds one excitation to the system which is exchanged due to the coupling between the transmons and subsequently removed from the system by the other drive (or some permutation of these processes). As one specific example, the perturbation mediates the virtual process that takes the system through the sequence of states $\ket{00} \rightarrow \ket{10} \rightarrow \ket{01} \rightarrow \ket{00}$ which produces corrections to $E_{00}$ at third order. With a drive on just one of the transmons this cannot occur and the leading corrections to $\zeta$ due to the drive induced conditional Stark shift appear only at fourth order. In total there are 2, 6, 6, and 12 terms in the corrections $E^{(3)}_{00}$, $E^{(3)}_{01}$, $E^{(3)}_{10}$, and $E^{(3)}_{11}$ respectively. When all of these are taken into account the $ZZ$ rate (through third order) is found to be
\begin{equation}
    \zeta=\zeta^{(2)}+\zeta^{(3)}=2J^{2}\left(\frac{1}{\Delta-\eta_{\mathrm{t}}}-\frac{1}{\Delta+\eta_{\mathrm{c}}}\right)+\frac{8\eta_{\mathrm{t}}\eta_{\mathrm{c}}J\varepsilon_{\mathrm{t}}\varepsilon_{\mathrm{c}}\cos\varphi}{\Delta_{\mathrm{c}}\Delta_{\mathrm{t}}(\Delta_{\mathrm{c}}+\eta_{\mathrm{c}})(\Delta_{\mathrm{t}}+\eta_{\mathrm{t}})}.
\end{equation}

In practice we typically have $\varepsilon_{i} \gg J$ and as a result $\zeta^{(3)}>\zeta^{(2)}$, as we demonstrate experimentally by showing that the drive induced $ZZ$ ($\zeta^{(3)}$) can cancel out the static $ZZ$ ($\zeta^{(2)}$). This calculation also recovers the linear dependence of $\zeta$ on the drive amplitudes $\varepsilon_{i}$ when $\varepsilon_{i} \gg J$. Lastly, this result matches the approximately sinusoidal dependence of the $ZZ$ rate on the relative drive phase $\varphi$ which is observed experimentally. Deviations from sinusoidal dependence are dominantly due to microwave crosstalk, and to a much lesser degree, higher order perturbative corrections with $\cos2\varphi, \cos3\varphi,\dots$ dependence. The fourth order perturbative corrections were also calculated and it was found that $\zeta^{(4)} \approx 0.01 \zeta^{(3)}$ for the range of experimental parameters that were explored.

\section{Numerical Simulations}
\begin{figure}[!th]
    \centering
    \begin{tikzpicture}
        \node[anchor=north west,inner sep=0] (image) at (0.0,-0.1cm) {\includegraphics[width=14.0cm]{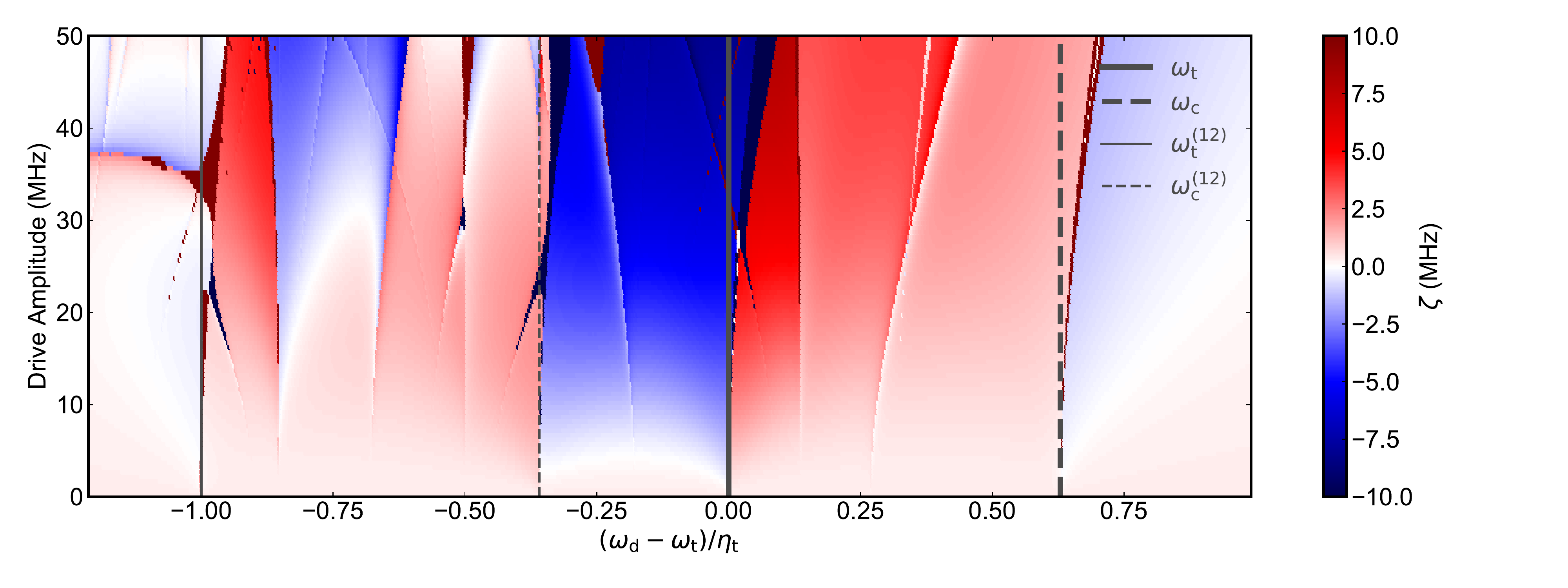}};
        \node[] at (0.0,-0.1) {\textbf{(a)}}; 
        \node[anchor=north west,inner sep=0] (image) at (0.0, -6cm)
        {\includegraphics[width=6.5cm]{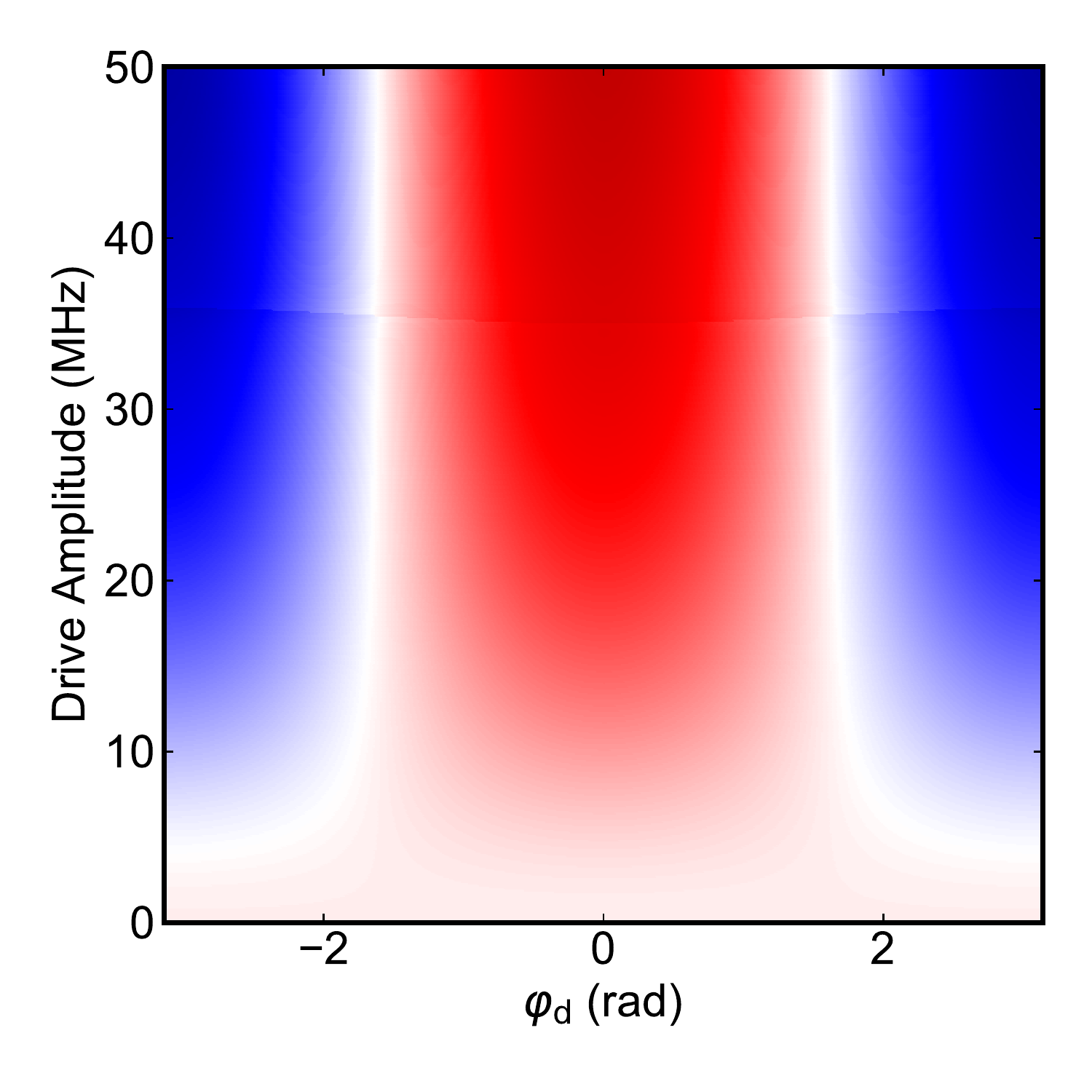}};
        \node[] at (0.0, -6.0cm) {\textbf{(b)}}; 
        \node[anchor=north west,inner sep=0] (image) at (6.2cm,-6.0cm)
        {\includegraphics[width=6.5cm]{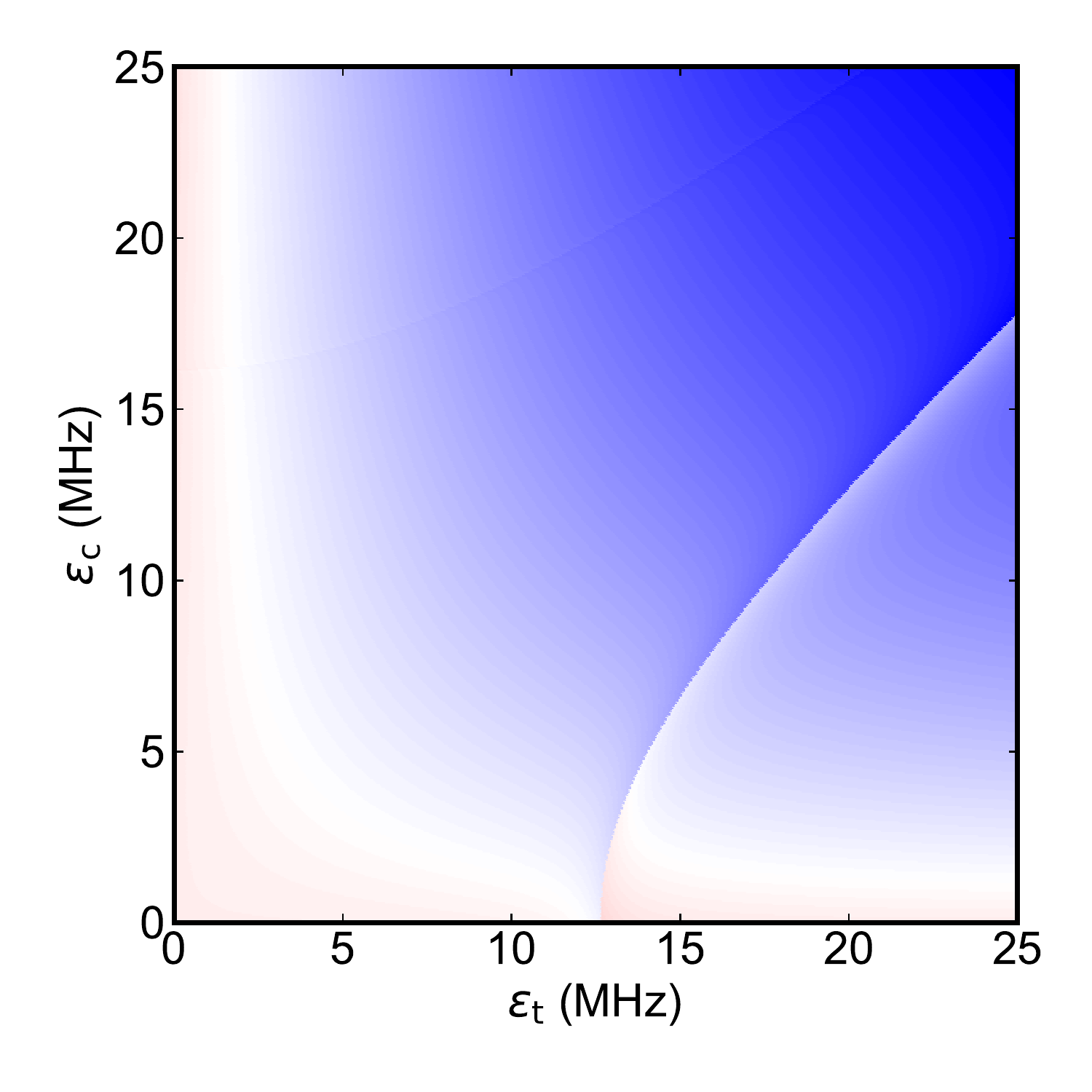}};
        \node[] at (6.7cm,-6.0cm) {\textbf{(c)}};
    \end{tikzpicture}
    \caption{\textbf{$ZZ$ Numerical Simulations} \textbf{(a)} $\zeta$ versus $\omega_\mathrm{d}$ and global drive amplitude $|\varepsilon|=|\varepsilon_\mathrm{t}|=|\varepsilon_\mathrm{c}|$, with $\varphi_\mathrm{d}=\pi$. Driving in the regions $\omega_\mathrm{c}^{(12)} < \omega_\mathrm{d} < \omega_\mathrm{t}$ and $\omega_\mathrm{t} < \omega_\mathrm{d} < \omega_\mathrm{c}$ show consistently larger areas of enhanced ZZ interaction. Driving below $\omega_\mathrm{c}^{(12)}$  or above $\omega_\mathrm{c}$, the $ZZ$ enhancement is reduced. There are many resonances visible in the simulation.  \textbf{(b)} $\zeta$ versus $\varphi_\mathrm{d}$ and $|\varepsilon|$, with $\omega_\mathrm{d} = \omega_\mathrm{t}-40\unit{MHz}$. \textbf{(c)} $\zeta$ versus $\varepsilon_\mathrm{c}$ and $\varepsilon_\mathrm{t}$, with $\varphi_\mathrm{d}=\pi$, $\omega_\mathrm{d} = \omega_\mathrm{t}-40\unit{MHz}$.}
    \label{fig:supp_zz_numerics}
\end{figure}

The simulation results presented in Figure 2 (b) and (c) of the main text were calculated numerically by diagonalizing the Hamiltonian in eq.~\ref{eq:Hsupp} using QuTiP~\cite{johansson_qutipqutip_2021}. Here, we present further simulation results in Figure~\ref{fig:supp_zz_numerics} showing how $\zeta$ behaves as a function of $\omega_\mathrm{d}$, $\varphi_\mathrm{d}$, $\varepsilon_\mathrm{c}$, and $\varepsilon_\mathrm{t}$. We include $7$ levels for each transmon in the calculation and use the parameters for the first presented pair: $\omega_{\mathrm{c}}/2\pi=5.845\unit{GHz}$, $\omega_{\mathrm{t}}/2\pi=5.690\unit{GHz}$, $\eta_{\mathrm{c}}/2\pi=-244.1\unit{MHz}$, $\eta_{\mathrm{t}}/2\pi=-247.1\unit{MHz}$, and $J = 3.45\unit{MHz}$. In Fig.~\ref{fig:supp_zz_numerics} (a), $\zeta$ is calculated for a range of drive frequencies and amplitudes. The regions of largest $\zeta$ are when $\omega_\mathrm{c}^{(12)} < \omega_\mathrm{d} < \omega_\mathrm{t}$, where $\omega_\mathrm{c}^{(12)} = \omega_\mathrm{c} + \eta_\mathrm{c}$, and $\omega_\mathrm{t} < \omega_\mathrm{d} < \omega_\mathrm{c}$. There are also many sharp features associated with the drive interacting resonantly with higher levels of the transmons. Fig.~\ref{fig:supp_zz_numerics} (b) shows the dependence of $\zeta$ on $\varphi_\mathrm{d}$ and $|\varepsilon|$ with $\omega_\mathrm{d} = \omega_\mathrm{t}-40\unit{MHz}$. The sinusoidal dependence is visible, as well as a resonance around $|\varepsilon|=37\unit{MHz}$. In Fig.~\ref{fig:supp_zz_numerics} (c) we sweep the amplitudes independently at the same drive frequency as before, keeping constant $\varphi_{\mathrm{d}}=\pi$.  The dependence on each amplitude is symmetric, except for the resonance observed for $\varepsilon_\mathrm{t} > 12.5\unit{MHz}$.

\section{Local $Z$ Gate Calibration}

\begin{figure}
    \centering
    \begin{tikzpicture}
        \node[] at (0.0,0.0cm) {\textbf{(a)}}; 
        \node[anchor=north west,inner sep=0] (image) at (0.0cm,0.0cm) {\includegraphics[width=5cm]{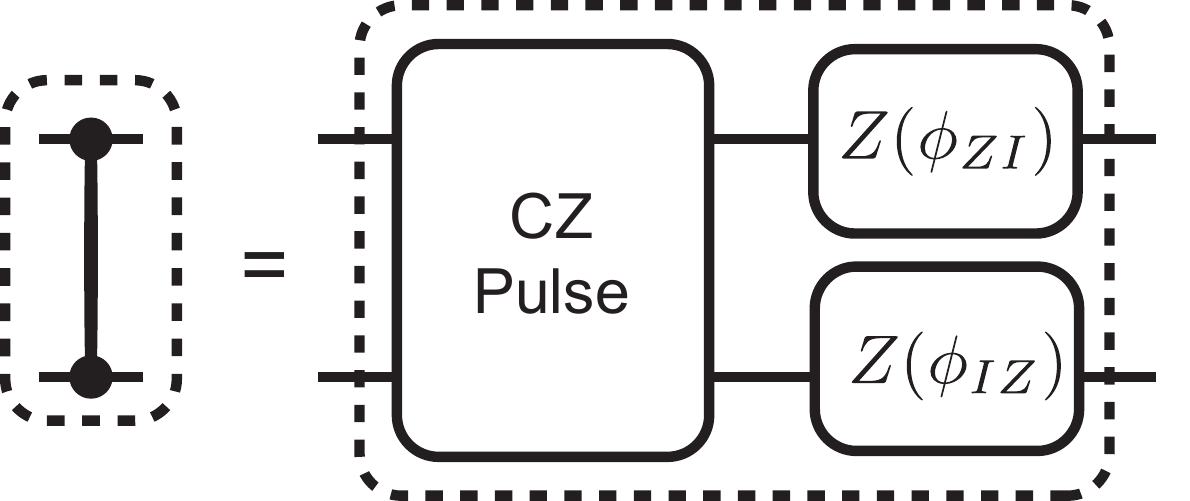}};
        \node[] at (5.3cm,0.0cm) {\textbf{(b)}}; 
        \node[anchor=north west,inner sep=0] (image) at (5.5cm,0.0cm) {\includegraphics[width=5cm]{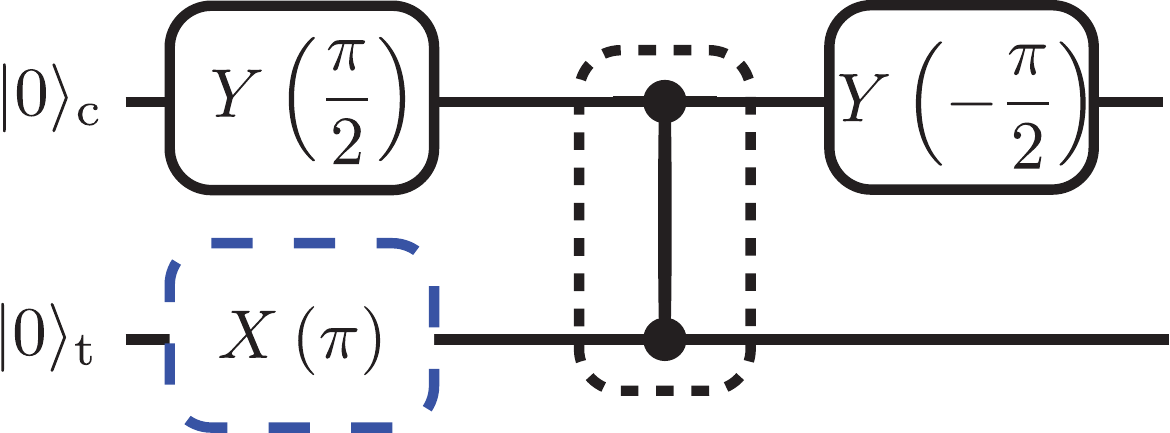}};
        
        \node[] at (0.0,-3cm) {\textbf{(c)}}; 
        \node[anchor=north west,inner sep=0] (image) at (0.0,-3.2cm) {\includegraphics[width=10cm]{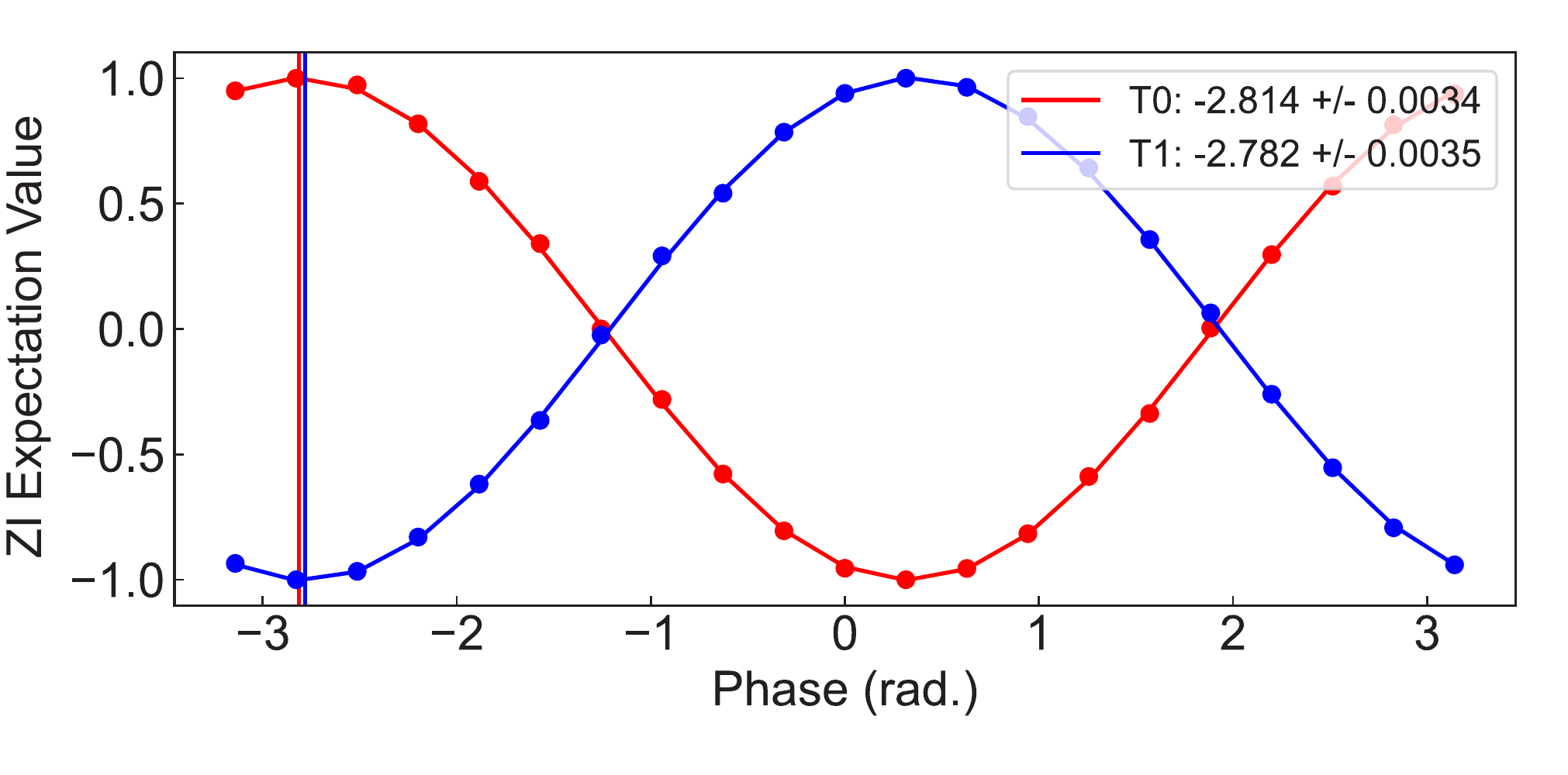}};
    \end{tikzpicture}
    \caption{\textbf{Local $Z$ Gate Calibration} \textbf{(a)} Quantum circuit for compiling a CZ gate using the entangling CZ pulse and local Z corrections. \textbf{(b)} Local phase calibration experiment. The control qubit is prepared along the x-axis of the Bloch sphere $\ket{+}=\frac{1}{\sqrt{2}}\left(\ket{0} + \ket{1}\right)$, and measured in the x-basis via $Y\left(\frac{\pi}{2}\right)$, $Y\left(-\frac{\pi}{2}\right)$ gates, respectively. \textbf{(c)} The local phase $\phi_{ZI}$ is swept while measuring $\left<ZI\right>$ for both target preparation states $\ket{0}$ (T0) and $\ket{1}$ (T1). The value of $\phi_{ZI}$ is calibrated when each input state is mapped to the correct output state, i.e., $\ket{+,0}\rightarrow \ket{0,0}$ and $\ket{+,1}\rightarrow \ket{11}$. This is analogously done for the target qubit local phase $\phi_{IZ}$.}
    \label{fig:supp_local_z}
\end{figure}

Here we outline how the local terms of the CZ gate are calibrated. After calibrating the $ZZ$ term of the Hamiltonian to the maximally entangling angle of $\frac{\pi}{2}$, the operation of the pulse on the qubits is of the form $\exp\left(-\frac{i}{2}\left(\alpha IZ + \beta ZI + \frac{\pi}{2}ZZ\right)\right)$. To correct for phases $\alpha$ and $\beta$ such that the phases on each qubit are aligned to realize a CZ gate, we apply virtual $Z$ (VZ) gates~\cite{mckay_efficient_2017} with angles $\phi_{ZI}$ and $\phi_{IZ}$ after the entangling pulse, as illustrated in Fig~\ref{fig:supp_local_z} (a). With the VZ gates, the combined unitary applied to the qubits is $\exp\left(-\frac{i}{2}\left(\left(\alpha + \phi_{IZ}\right) IZ + \left(\beta + \phi_{ZI}\right) ZI +  \frac{\pi}{2}ZZ\right)\right)$. To calibrate $\phi_{ZI}$, we prepare the control qubit along the x-axis of the Bloch sphere $\ket{+}=\frac{1}{\sqrt{2}}\left(\ket{0} + \ket{1}\right)$, apply the CZ gate circuit, and measure along the x-axis, as shown in Fig.~\ref{fig:supp_local_z} (b). We sweep $\phi_{ZI}$ and measure $\left<ZI\right>$ when the target qubit is in $\ket{0}$ and $\ket{1}$, as shown in Figure~\ref{fig:supp_local_z} (c). The value of $\phi_{ZI}$ is calibrated when each input state is mapped to the correct output state, i.e., $\ket{+,0}\rightarrow \ket{0,0}$ and $\ket{+,1}\rightarrow \ket{11}$.  We fit the expectation values for both target preparations to and set the calibrated value to the average of the two fit results. This experiment is analogously applied for $\phi_{IZ}$, exchanging the roles of the control and target qubit in the experiment.

\section{Coherence Dependence on Drive Amplitude}

\begin{figure}
    \centering
    \begin{tikzpicture}
         
        \node[anchor=north west,inner sep=0] (image) at (0.0cm,0.0cm) {\includegraphics[width=5cm]{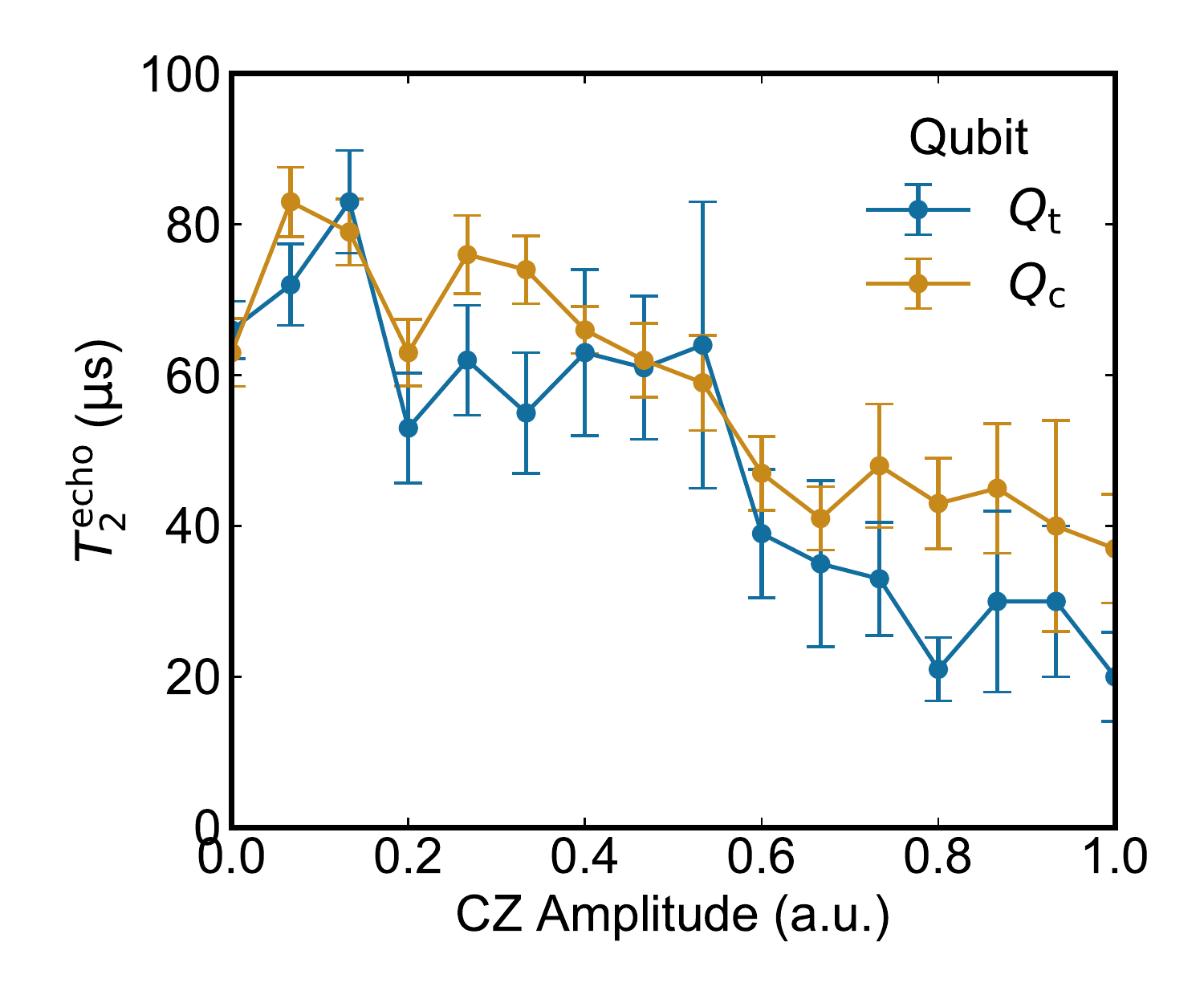}};
        \node[] at (0.0,0.0cm) {\textbf{(a)}};
        
        \node[anchor=north west,inner sep=0] (image) at (5.2cm,0.0cm) {\includegraphics[width=5cm]{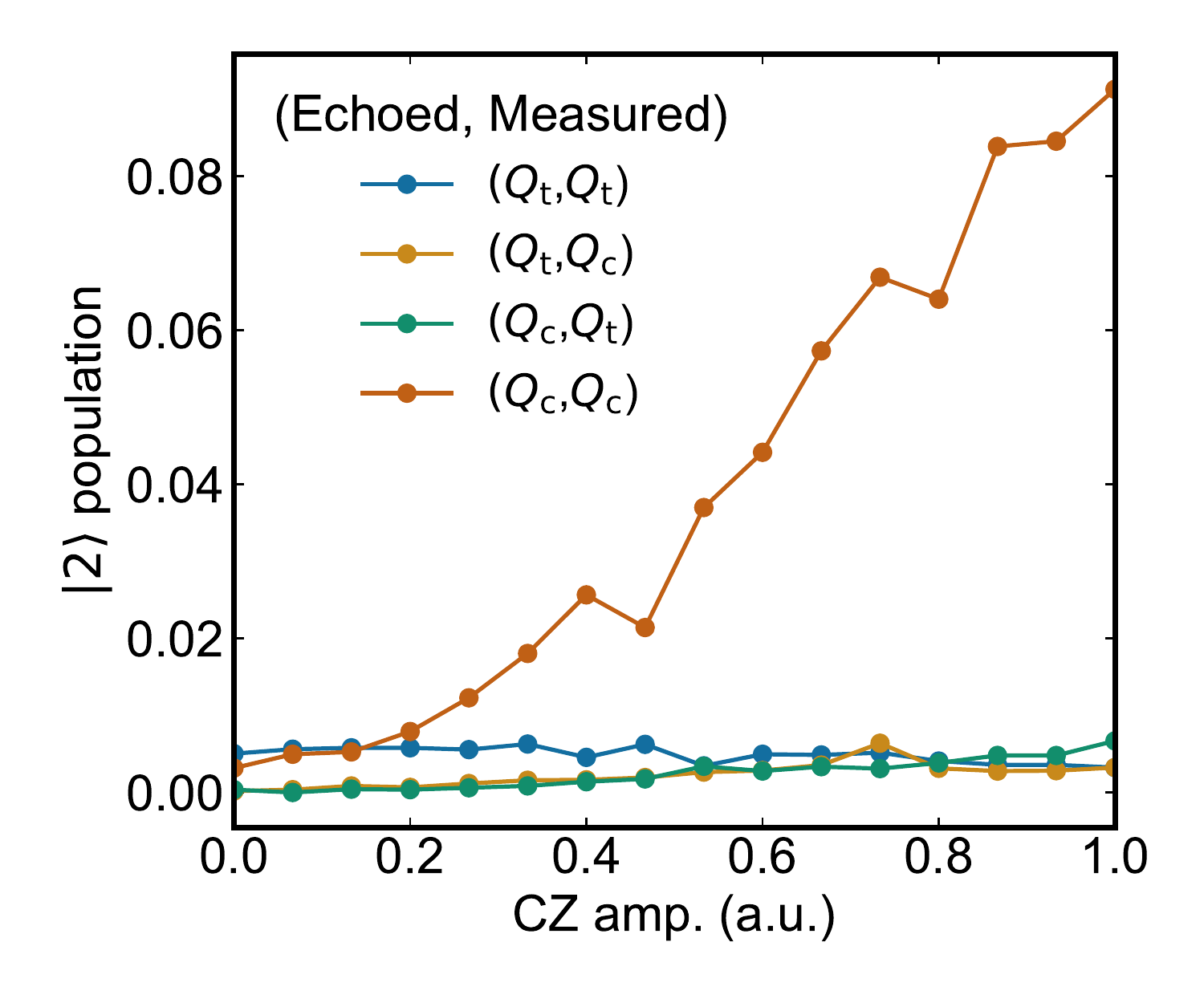}};
        \node[] at (5.2cm,0.0cm) {\textbf{(b)}}; 
        
        \node[anchor=north west,inner sep=0] (image) at (10.4cm,0.0) {\includegraphics[width=5cm]{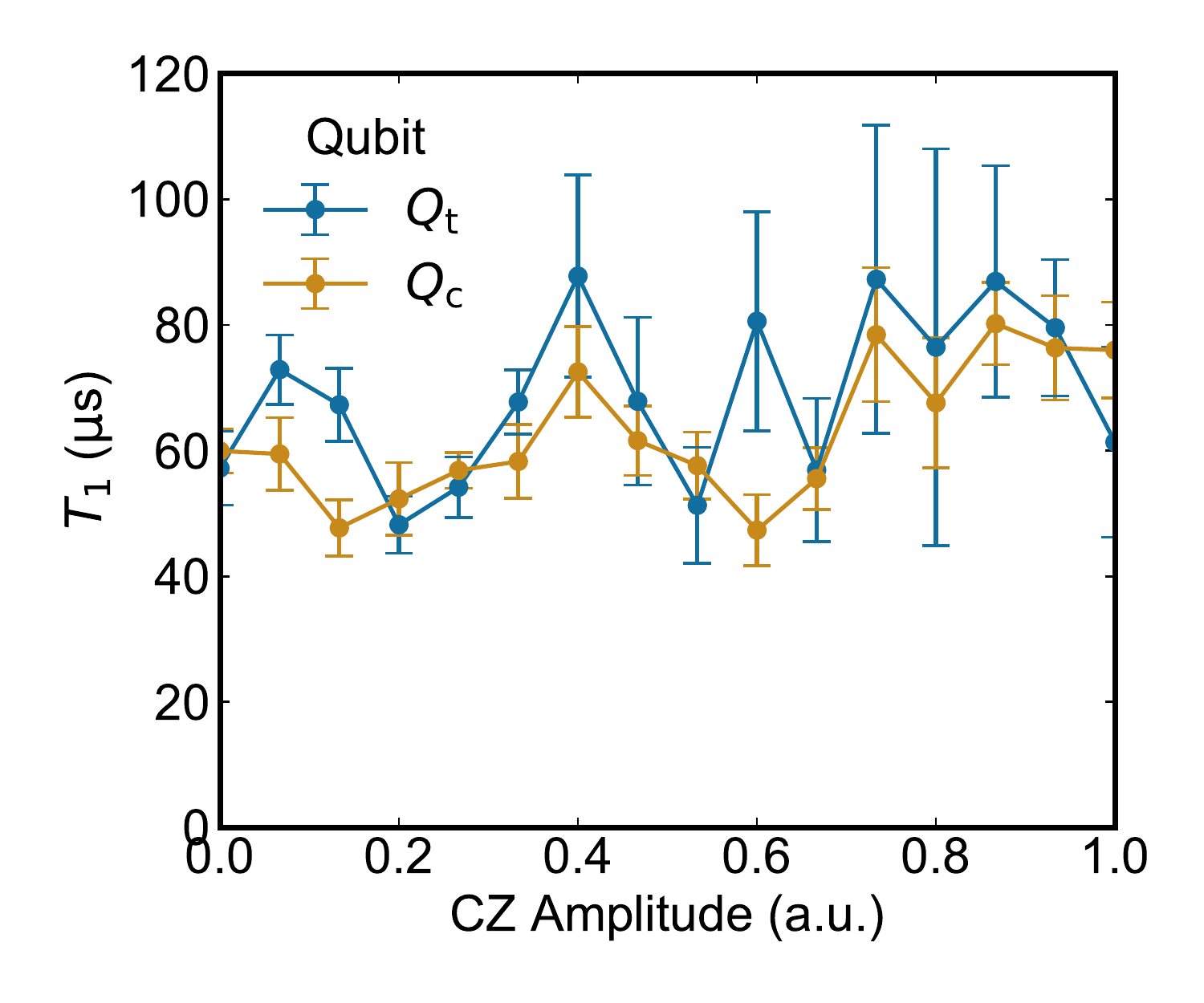}};
        \node[] at (10.4cm,0.0cm) {\textbf{(c)}}; 
    \end{tikzpicture}
    \caption{\textbf{Coherence versus Drive Amplitude}. \textbf{(a)} Measurements of $T_2^{\mathrm{echo}}$ versus CZ drive amplitude for both qubits used in the gate in the main text. Error bars are uncertainties in the fit. The coherence of the sample is reduced at larger CZ pulse amplitudes.\textbf{(b)} Average $\ket{2}$-state population during the echo experiments, versus CZ drive amplitude. We observe largest leakage to $\ket{2}$ for $Q_\mathrm{c}$ when measuring $Q_\mathrm{c}$, suggesting leakage from $\ket{1}\rightarrow\ket{2}$ for $Q_\mathrm{c}$ is the dominant source of leakage for this interaction. \textbf{(c)} Qubit lifetimes $T_1$ as a function of CZ pulse amplitude. No visible trend is observed. }
    \label{fig:techo_vs_cz_amp}
\end{figure}

To investigate whether the CZ drive generated additional sources of incoherent noise, we measured $T_2^{\mathrm{echo}}$ when the CZ drive is turned on, as a function of CZ pulse amplitude. The results are shown in Figure~\ref{fig:techo_vs_cz_amp} (a). We see that the echo time decreases with increasing CZ pulse amplitude, consistent with the discrepancy between measured stochastic errors of the gate and what is predicted by coherence estimates, described in the main text. We also monitored the average $\ket{2}$-state population during in these experiments (see Figure~\ref{fig:techo_vs_cz_amp} (b)). The largest leakage is observed $Q_\mathrm{c}$ when performing the echo experiment on $Q_\mathrm{c}$, indicating off-resonant driving of the $\ket{1}\rightarrow\ket{2}$ transition of $Q_\mathrm{c}$. Finally, in Figure~\ref{fig:techo_vs_cz_amp} (c) we show $T_1$ versus CZ pulse amplitude, where little trend is observed, as expected. The first 10 data points (about $5\unit{\mu s}$) were omitted from the time series data when fitting, due to off-resonant driving of the transmons at high amplitudes which caused deviation from exponential decay of the qubit populations.

\section{Leakage Randomized Benchmarking}

To estimate leakage-per-gate, we perform interleaved randomized benchmarking while resolving the $\ket{2}$-state for both transmons. Imperfect $\ket{2}$-state readout fidelity results in an underestimate of the leakage rate in a way approximately proportional to the $\ket{2}$-state readout infidelity~(see \cite{chen_measuring_2016} supplement). For our LRB experiment, the readout fidelities for the three states are $p_\mathrm{c}^{\ket{0}} = 0.996$, $p_\mathrm{c}^{\ket{1}} = 0.984$, $p_\mathrm{c}^{\ket{2}} = 0.975$ for the control transmon, and $p_\mathrm{t}^{\ket{0}} = 0.996$, $p_\mathrm{t}^{\ket{1}} = 0.984$, $p_\mathrm{t}^{\ket{2}} = 0.975$ for the target transmon, suggesting that the leakage rates we measure are a few percent smaller than the actual leakage rates. We fit the observed $\ket{2}$ state populations as a function of Clifford sequence length $m$ from both the referenced and interleaved IRB experiments to the exponential model 
\begin{align}
P_{\ket{2}} &= B - A e^{-\Gamma m} \\
\Gamma &= \gamma_\uparrow + \gamma_\downarrow \\
B &= \gamma_\uparrow/\Gamma,
\end{align}
where $\gamma_\uparrow$ is the leakage rate and  $\gamma_\downarrow$ is the seepage rate~\cite{chen_measuring_2016,wood_quantification_2018}. Fitting both interleaved and reference datasets to this model, we extract the leakage-per-gate as $\gamma_\uparrow^{\mathrm{CZ}}=\gamma_\uparrow^{\mathrm{IRB}} - \gamma_\uparrow^{\mathrm{RB}}$~\cite{krinner_demonstration_2020}.

\clearpage
\bibliographystyle{apsrev4-1}
\bibliography{cz_refs_supp}